\documentclass[prd,preprintnumbers,pacs,12pt,nofootinbib]{revtex4-1}

\pdfoutput=1

\usepackage{amsmath}
\usepackage{amssymb}
\usepackage{overpic,color}
\usepackage{feynmf}
\usepackage{epsfig}
\usepackage{graphicx} % Ricorda sempre: la \label nelle figure va messa DOPO la caption, o il numero di riferimento viene sbagliato...
\usepackage{epstopdf} % Ricorda sempre: questo package, che mac os - specifico, va messo DOPO graphics o graphicx

\usepackage{bm}
\usepackage{amsmath}
\usepackage{amssymb}

\usepackage[latin1]{inputenc}
\usepackage{amscd}
\usepackage{latexsym}
\begin{document}

\newcommand{\bd}{\begin{defn}}
\newcommand{\ed}{\end{defn}}
\newcommand{\bt}{\begin{teo}}
\newcommand{\et}{\end{teo}}
\newcommand{\bl}{\begin{lem}}
\newcommand{\el}{\end{lem}}
\newcommand{\bc}{\begin{cor}}
\newcommand{\ec}{\end{cor}}
\newcommand{\bp}{\begin{prop}}
\newcommand{\ep}{\end{prop}}
\newcommand{\bo}{\begin{oss}}
\newcommand{\eo}{\end{oss}}

\newcommand{\nn}{\nonumber}
\newcommand{\bea}{\begin{eqnarray}}
\newcommand{\eea}{\end{eqnarray}}

 \newcommand{\de}{\partial}
\newcommand{\fre}{\rightarrow}
\newcommand{\Fre}{\Rightarrow}
\newcommand{\la}{\mathcal L}
\newcommand{\con}{\dag}
\newcommand{\bi}{\begin{itemize}}
\newcommand{\ei}{\end{itemize}}
\newcommand{\be}{\begin{enumerate}}
\newcommand{\ee}{\end{enumerate}}
\newcommand{\spa}{\,\,\,\,\,\,\,\,\,\,\,}
\newcommand{\De}{\tilde{\mathcal D}}
\newcommand{\Du}{\mathcal D}
\newcommand{\Span}{\text{Span}}
\newcommand{\C}{\mathbb C}
\newcommand{\R}{\mathbb R}
\newcommand{\K}{\mathbb K}
\newcommand{\Z}{\mathbb Z}
\newcommand{\Q}{\mathbb Q}
\newcommand{\N}{\mathbb N}
\newcommand{\D}{\mathbb D}
\newcommand{\dd}{\displaystyle}
\newcommand{\calA} {{{\mathcal A}}}
\newcommand{\OO}{{\cal O}}

%\tableofcontents
\title{Effective chiral Lagrangians for new vector bosons to $\OO(p^4)$ order}

\author{Francesco Bernardini, Francesco Coradeschi and Daniele Dominici}
\affiliation{Department of Physics, University of Florence, and
 INFN,  Via Sansone 1, 50019 Sesto F., (FI), Italy}

\date{\today}

\begin{abstract}
\noindent
If the SM Higgs boson does not exist, electroweak symmetry breaking may be realized via a strong interaction with a typical scale $\Lambda > 1$ TeV.
Resonances from the strong sector may help to unitarize $WW$ scattering, which becomes strong in the absence of an Higgs field, and could be detected at the LHC.

In this paper we describe such a scenario, in the minimal case in which only one new vector resonance is present, via a chiral $SU(2) \otimes SU(2)/SU(2)$ lagrangian also including all possible invariant terms up to $\OO(p^4)$ order (assuming a parity symmetry in the strong sector). The $\OO(p^4)$ invariants are not usually taken into account in similar studies in the literature; however, they have been shown to be potentially important, at least, to reconcile this kind of scheme with electroweak precision tests. Here we use the $\OO(p^4)$ lagrangian to study the scattering amplitudes for the $\pi\pi$ sector, investigating the partial wave unitarity properties of the model and its cut-off energy scale. We obtain constraints on the parameter space and compare our result to the one obtained with just the $\OO(p^2)$ lagrangian, finding that the contribution of the new operators is indeed significant.
\end{abstract}

\maketitle

\section {Introduction}

One of the main purposes of the LHC is the understanding of the mechanism behind electroweak symmetry breaking. Finding an SM-like Higgs boson would not provide a complete answer to this question, since the Higgs mechanism, as implemented in the SM, suffers from the well-known hierarchy problem: the Higgs boson mass parameter, and thus the electroweak scale, is unstable against quantum corrections. So, if the Higgs is found, we will still need to understand what is the mechanism that stabilizes its mass.

On the other hand, the possibility that neither the SM Higgs nor any other scalar resonance with similar properties exist is still open. Higgsless models have long since been proposed as one possible realization of the broader scenario in which the electroweak symmetry breaking is brought about by a new strong interaction, which should reside somewhere around the 1 TeV scale. Building fully explicit models along these lines is not easy, since theories in the nonperturbative regime are notoriously difficult to treat. However, the properties of a possible new strong sector can be explored by using model-independent approaches; in particular, chiral Lagrangian techniques, already successfully used for low energy QCD \cite{Weinberg:1978kz,Gasser:1983yg,Gasser:1984gg}, have also been applied to the study strong electroweak symmetry breaking \cite{Appelquist:1980vg,Longhitano:1980iz,Casalbuoni:1985kq,Casalbuoni:1986vq,Casalbuoni:1988xm}.

In the SM without a Higgs boson, the scattering amplitudes of longitudinally polarized $W$ and $Z$ bosons typically violate unitarity around $\sim$ 1.7 TeV \cite{Lee:1977eg}. The exchange of light Higgs bosons, however, cancels the unitarity-violating terms and ensures perturbativity of the theory up to high scales. By contrast, effective Higgsless models based on chiral Lagrangians are nonrenormalizable, and unitarity is necessarily violated at some cut-off scale. However, the exchange of resonances from the strong sector should help to mitigate the unitarity behaviour of the theory, postponing the violation scale with respected to the SM case.

A renewed interest in chiral Lagrangians has been stimulated by the discovery of theories in extra dimensions \cite{Antoniadis:1998ig,ArkaniHamed:1998rs,Randall:1999ee,Randall:1999vf} which postulate a scale of extra dimensions  related to the Fermi scale, and are related by the AdS/CFT correspondence \cite{Maldacena:1997re,Gubser:1998bc,Witten:1998qj} to strongly interacting theories in four dimensions. Gauge models formulated in five dimensions allow for new descriptions of electroweak breaking with the Higgs \cite{Csaki:2002gy,Agashe:2003zs} or without \cite{Csaki:2003dt,Csaki:2003sh,Nomura:2003du} making use of boundary conditions at the ends of the fifth dimension. In the latter case, it has been shown that the exchange of the heavier KK gauge bosons does indeed cancel the dominant unitarity-violating terms \cite{SekharChivukula:2001hz,Csaki:2003dt,Ohl:2003dp,Papucci:2004ip,Muck:2004br,Falkowski:2007iv}, pushing the scale of unitarity violation forward in the TeV range.

The five-dimensional models are closely related to, and have helped understand, the ones obtained via four-dimensional chiral models: effective low energy chiral Lagrangians in four dimensions can be obtained from extra-dimensional theories via the deconstruction technique \cite{ArkaniHamed:2001ca,ArkaniHamed:2001nc,Hill:2000mu,Cheng:2001vd,Abe:2002rj,Falkowski:2002cm,Randall:2002qr,Son:2003et,deBlas:2006fz}, providing a correspondence at low energies between effective chiral Lagrangians with replicated 4D gauge symmetries $G$ and theories with a 5D gauge symmetry $G$ on a lattice. Simplest models of chiral Lagrangians with three or four sites have been studied by considering chiral effective Lagrangian to second order in the derivative expansion \cite{Casalbuoni:1985kq,Casalbuoni:1986vq,Casalbuoni:1988xm,SekharChivukula:2006cg,Foadi:2007ue,Barbieri:2008cc,Barbieri:2009tx,Accomando:2008jh,Falkowski:2011ua}.

Loop corrections have been recently shown to be important for reconciling these schemes with electroweak precision data \cite{Dawson:2007yk,Barbieri:2008cc,Abe:2008hb}. Therefore it is necessary to develop chiral Lagrangians to fourth order in the derivative expansion since fourth order vertex operators in the effective Lagrangian a priori generate contributions which are of the same order of the one loop expansion terms. Some preliminary work in this direction can be found in \cite{Tanabashi:1993sr} while in a recent paper \cite{Contino:2011np} the general chiral Lagrangian for the composite Higgs model based on the $SO(5)/SO(4)$ coset to $\OO(p^4)$ has been considered.

In this work, we will obtain the $\OO(p^4)$ chiral Lagrangian in the simplest effective Higgsless model, in which a single vector resonance from the strong sector is included. We will then use the result to evaluate high-energy $W$ and $Z$ scattering to study unitarity violation, and determine to which degree the single resonance can push up the cut-off scale of the model, putting special emphasis on the role of the new $\OO(p^4)$ operators.

The outline of the paper is as follows. In Section \ref{hgs} we briefly review the formalism of hidden gauge symmetry which we will use to include the new resonance. In sec. \ref{model} we present the actual model, building its Lagrangian up to $\OO(p^4)$. In Section \ref{unitconstr} we consider amplitudes for $V^{SM}_L V^{SM}_L$ scattering (with $V^{SM} = W, \, Z$) and use them to obtain the unitarity limit as a function of the model parameter space. Conclusions and final comments are given in Section \ref{conclusions}. A short sketch of the proof for discovering the independent quadrilinear $\OO(p^4)$ operators is performed in appendix \ref{OpTraces}. Feynman rules relevant for $V^{SM}_L V^{SM}_L$ scattering are given in appendix \ref{appB}.

\section{The hidden gauge symmetry formalism}
\label{hgs}
In this section we briefly review the \emph{hidden gauge symmetry} approach which
is one of the methods which are routinely used in the literature to include new vectors in non linear $\sigma$-models \cite{Balachandran:1978pk,Bando:1984ej,Bando:1984pw,Bando:1985rf,Bando:1987br}. The Lagrangian for the model, which describes the low energy limit of a theory with a global symmetry $G$ spontaneously breaking to a subgroup $H$, is built by using the Maurer-Cartan form associated to a field variable $g$ which lives in an unitary matrix representation of $G$, that is $\omega_\mu (g) = g^\dag\de_\mu g$. The field $g$ transforms as
\begin{equation}
g(x) \to g_0 \, g (x) \, h(x), \quad g \in G, \ h \in H_{loc}
\end{equation}
under the action of $G \otimes H_{loc}$, where $H_{loc}$ is a \emph{local} copy of the subgroup $H$. The Maurer-Cartan form decomposes as
\begin{equation}
\omega_\mu = \omega^\bot_\mu + \omega^\parallel_\mu
\end{equation} 
where
$\omega^\parallel_\mu$ is along the unbroken subgroup $H$ and $\omega^\bot_\mu$ along $G/H$, and transforms as
\begin{equation}
\begin{array}{l}
\dd\omega^\parallel_\mu(g_0\,g\,h)=h^\dag\omega^\parallel_\mu(g)h +h^\dag\de_\mu h\\
\dd\omega^\bot_\mu(g_0\,g\,h)=h^\dag\omega^\bot_\mu(g)h.
\end{array}\label{eq2}
\end{equation} 
By introducing the generators
$H_\nu$ of $Lie[H]$ and $X_i \in G/H$ with the usual normalization,
\begin{equation}
Tr[S_a S_b]=\frac12\delta_{ab}, \quad S = H, X,
\label{norm}
\end{equation}
we get
\begin{equation}
\omega^\parallel_\mu(g)=2H_\rho Tr[H_\rho\omega_\mu(g)],\spa \omega^\bot_\mu(g)=2X_i Tr[X_i\omega_\mu(g)].
\end{equation}
The transformations \eqref{eq2} suggest the introduction of a gauge field $V_\mu$ transforming as
\begin{equation}
V_\mu\fre h^\dag V_\mu h+h^\dag \de_\mu h
\end{equation}
so that the combinations $\omega^\bot$ and $\omega^\parallel - V$ both transform covariantly under $H$.
For generic $G$ and $H$, the most general Lagrangian up to order $\OO(p^2)$ is then built as
\begin{equation*}
\la=-f^2\left[I_1^{(2)}+\alpha I_2^{(2)} \right]
\end{equation*}
with
\begin{equation}
I_1^{(2)}=Tr[{\omega^\bot}^2],\spa I_2^{(2)}=Tr[(\omega^\parallel-V)^2]
\label{I1I2}
\end{equation}
and $\alpha$ and $f$ arbitrary parameters.

If the coset $G/H$ is a symmetric space, that is commutation relations are of the form
\begin{equation}
[H_\mu,H_\nu]=if_{\mu\nu\lambda}H_\lambda,\spa[X_i,X_j]=if_{ij\mu} H_\mu,\spa [H_\mu,X_j]=if_{\mu ij}X_j
\end{equation}
(equivalently, if $H$ is a normal subgroup of $G$), it is possible to introduce a discrete symmetry $R$ such that
\begin{equation}
R(H_\mu) = H_\mu, \quad R(X_i) = -X_i
\label{auto}
\end{equation}
that may be useful in classifying invariants.

If no other ingredient is added, then the field $V$ is an auxiliary field that can be removed using the equations of motion; this leads to $I_2^{(2)} = 0$ and recovers the CCWZ formulation \cite{Coleman:1969sm,Callan:1969sn} of the low energy $G \to H$ spontaneous breaking dynamics. However, $V$ can be rendered dynamical by the addition of a kinetic term $F_{\mu\nu}(V) F^{\mu\nu} (V)$ (note that the field strength $F_{\mu\nu}(V)$ is covariant under the action of $H_{loc}$, just as $\omega_\mu^\bot$ and $(\omega_\mu^\parallel-V)$, so that $F_{\mu\nu}(V) F^{\mu\nu} (V)$ is an invariant, in particular of $\OO(p^4)$). The field $V$ acquires mass by the Higgs mechanism, eating $dim(H)$ of the $(dim(G)-dim(H))$ scalar degrees of freedom contained in $\omega_\mu$, and can be interpreted as a resonance of the strong interaction driving the $G \to H$ spontaneous breaking. The above technique has been applied both to QCD (\cite{Bando:1984ej,Bando:1985rf}), by identifying the new vectors as the $\rho$ mesons, and to the electroweak strong symmetry breaking (\cite{Casalbuoni:1985kq,Casalbuoni:1986vq}), in the latter case by choosing $G=SU(2)_L \times SU(2)_R$, $H = SU(2)_D$ and postulating the existence of new vector resonances in the electroweak sector. The construction can also be generalized, introducing more local copies of the vacuum symmetry $H$ which give rise to more vector (or axial vector) resonances \cite{Casalbuoni:1988xm}. 

Let us end this section by noting that other techniques have been used in the literature to describe vector resonances from a strong sector, including the vector as an adjoint matter field \cite{Ecker:1989yg} or as an antisymmetric vector field \cite{Ecker:1988te}. However, as it has been shown in several papers \cite{Ecker:1989yg,Pallante:1992qe,Bijnens:1995ii,Tanabashi:1995nz}, all these descriptions are equivalent at the level of on-shell amplitudes, so choosing one over the another is to some extent arbitrary, and should be based on the convenience of using a given formalism for the specific calculation one has in mind, rather than on physical grounds.

\section{The $SU(2)_L \otimes SU(2)_R$ chiral model}
\label{model}

In the following, we will be interested in the minimal scenario relevant for strong electroweak symmetry breaking, in which just a single new vector resonance is introduced. We will assume that the strong sector has a $G=SU(2)_L\otimes SU(2)_R$ symmetry, which is spontaneously broken to $H=SU(2)_V$, which is identified with the SM custodial symmetry.

We will make use of the formalism of sec. \ref{hgs}, and represent the field variable $g$ as the direct sum $L \oplus R$, where $L,R\in SU(2)_{L,R}$; the transformation under $G \otimes H_{loc}$ is, explicitly:
\begin{equation}
g=\left( \begin{array}{cc}
L & 0\\
0 & R
	\end{array} \right)\fre \left(
	\begin{array}{cc}
g_L Lh & 0\\
0 & g_R Rh
	\end{array} \right),
\label{2.21}
\end{equation} 
with $g_{L(R)}\in SU(2)_{L(R)}, \, h(x) \in SU(2)_V$. Symmetry is broken by requiring the VEV for $L,R$ fields to be equal to 1:
\begin{equation}
\left\langle L(x)\right\rangle =\left\langle R(x)\right\rangle =1;
\label{2.2}
\end{equation}
the diagonal subgroup is generated by $H^a=\tau_L^a + \tau_R^a$ with $\tau_L^\alpha=\dfrac{\sigma^\alpha}{2} \oplus 0_{2\times2}$, $\tau_R^\alpha=0_{2\times2} \oplus \dfrac{\sigma^\alpha}{2}$; the $\sigma^\alpha$ are the Pauli matrices. To complete $H^a$ to a $Lie[SU(2)_L\otimes SU(2)_R]$ basis, we choose as generators:
\begin{equation}
X^a=\tau_L^a -\tau_R^a;
\label{2.15}
\end{equation}
it is easy to verify that the $H^a$ and $X^a$ obey the orthonormality condition \eqref{norm}.
The commutation relations of the $SU(2)_L\otimes SU(2)_R$ algebra are, in terms of the $X^a$ and the $H^b$:
\begin{equation}
\dd [H^a,H^b]=i\epsilon^{abc}H^c, \quad
\dd [X^a,H^b]=i\epsilon^{abc}X^c, \quad
\dd [X^a,X^b]=i\epsilon^{abc}H^c;
\label{SU2comms}
\end{equation}
this tells use that $G/H$ is a symmetric space. The discrete symmetry of eq. \eqref{auto} can be in this case realized as the transformation $P_{LR}$ that exchanges the order of the blocks in $g$, $P_{LR}: L \leftrightarrow R$. If, as in the case of QCD, the global symmetry $SU(2)_L \otimes SU(2)_R$ is a flavour symmetry of chiral fermions, $P_{LR}$ is related to the ordinary parity $P$ as $P = P_0 + P_{LR}$, where $P_0$ is the spatial parity $P_0: (t, \vec{x}) \to (t, - \vec{x})$. In the following, we are going to assume, for the sake of simplicity, that the strong sector is invariant under $P_{LR}$ and $P_0$ separately (our choice of terminology here is borrowed from \cite{Contino:2011np}).

With the given notations, it is easy to find an explicit form for the Maurer-Cartan field. We have
\begin{equation}
\omega_\mu \equiv g^\dag\de_\mu g = L^\dag\de_\mu L\oplus R^\dag\de_\mu R,
\label{2.17}
\end{equation}
and by a change of basis we are able to express Cartan field on the basis $\{H^a,X^b\}$:
\begin{equation}
\begin{array}{l}
\dd \omega_\mu=g^\dag\de_\mu g=Tr[L^\dag\de_\mu L \tau_L^a] \tau_L^a + Tr[R^\dag\de_\mu R \tau_R^a] \tau_R^a \equiv\\
\dd =\frac12 (Tr[L^\dag\de_\mu L \tau_L^a] + Tr[R^\dag\de_\mu R \tau_R^a])H^a + \frac12 (Tr[L^\dag\de_\mu L \tau_L^a] - Tr[R^\dag\de_\mu R \tau_R^a]) X^a,
\end{array}
\label{2.19}
\end{equation}
which implies $\omega^\bot_\mu= Tr[g^\dag\de_\mu g X^a]X^a$ and $\omega^\parallel_\mu= Tr[g^\dag\de_\mu g H^a]H^a$. It is immediate to see that $\omega^\bot$ and $\omega^\parallel$ have well defined parity properties:
\begin{equation}
\label{2.30}
\omega^\bot_\mu \xrightarrow{P_{LR}\,} -\omega^\bot_\mu, \quad \omega^\parallel_\mu \xrightarrow{P_{LR}\,} \omega^\parallel_\mu.
\end{equation}
The $(SU(2) \otimes SU(2))/SU(2)_D$ coset has an additional symmetry with respect to the general case. The commutation relations \eqref{SU2comms} tells us that, in fact, the broken generators $X^a$ actually live in the adjoint representation of $SU(2)_D$. This means that the triplets $X^a$ and $H^a$ actually \emph{transform the same way} under the action of $H$. So, it makes sense to define the combination:
\begin{equation}
\hat{\omega}^\bot_\mu \equiv Tr[g^\dag\de_\mu g X^a]H^a,
\end{equation}
which has the following transformation properties under $H$ and $P_{LR}$:
\begin{equation}
\hat{\omega}^\bot_\mu \xrightarrow{H}  h^\dag \hat{\omega}^\bot_\mu h, \quad \hat{\omega}^\bot_\mu \xrightarrow{P_{LR}\,}  -\hat{\omega}^\bot_\mu.
\end{equation}
Given the properties of the generators $X^a$ and $H^a$, it is sufficient to just use $\hat{\omega}^\bot_\mu$ to build invariants; the use of $\omega^\bot_\mu$ does not lead to any independent contribution. This is shown explicitly in appendix \ref{OpTraces}.

The formalism described so far only accounts for the degrees of freedom that arise from the composite sector, that is the vector resonance $V$ and the three Goldstone scalars that provide the longitudinal components for the $W$ and $Z$ bosons. The weak interactions themselves have to be added at a later stage; this can be done by gauging a $SU(2) \otimes U(1)$ subgroup of the global symmetry $G$. Correspondingly, the derivatives in the Cartan field $\omega_\mu$ have to be generalized to covariant derivatives:
\begin{equation}
\de_\mu L \to \de_\mu L + i g W_\mu^a \frac{\sigma^a}{2} L, \quad \de_\mu R \to \de_\mu R - i g' B_\mu R \frac{\sigma^3}{2}.
\end{equation}

We are now ready to construct the effective Lagrangian up to $\OO(p^4)$. The building blocks for the construction of the invariants are $\hat{\omega}^\bot_\mu$, $\hat{\omega}_\mu^\parallel \equiv \omega_\mu^\parallel-V_\mu$, $F_{\mu\nu}(V)$ and all the operators we can obtain from these by means of the application of the covariant derivatives $\Du_\mu=\partial_\mu -V_\mu$ and  $\De_\mu=\partial_\mu -\omega_\mu^{\parallel}$; however, it is possible to show, proceeding as in \cite{Appelquist:1980vg}, that we can neglect all bilinear and trilinear invariants involving covariant derivatives, because using the equations of motion or convenient operator identities, they can be expressed in terms of simpler invariants. Also, when we gauge $SU(2) \otimes U(1)$ to add the weak interactions, two more operators, containing the kinetic terms for $W_\mu^a$ and $B_\mu$, can be added. These are $\hat{F}_{\mu\nu}(W)=Tr[L^\dag F_{\mu\nu}(W) L \, \tau^a]H^a$ and $\hat{F}_{\mu\nu}(B)=Tr[R^\dag F_{\mu\nu}(B) \sigma^3 R \, \tau^a]H^a$. The behaviour of these operators is similar to the one of $F_{\mu\nu}(V)$, as they are both triplets of $[SU(2)_V]_{loc}$. However, while $\hat{F}_{\mu\nu}(W)$ is $G$-invariant, $\hat{F}_{\mu\nu}(B)$ explicitly breaks $G$ to $SU(2) \otimes U(1)$, so that, when the scalars assume their VEV, only the $U(1)_{e.m.}$ subgroup remains unbroken, just as in the SM. Also, these terms automatically break $P_{LR}$. Coherently with our assumptions, the full Lagrangian will contain all the independent, local combinations of these ingredients that are both Lorentz- and $(G\otimes P_{LR})$-invariant in the limit in which no electroweak fields are considered, and Lorentz- and $SU(2) \otimes U(1)$ gauge-invariant when electroweak interactions are switched on.

The Lagrangian to order $\OO(p^2)$ just contains the two invariants $I_1^{(2)}$ and $I_2^{(2)}$ defined in eq. \eqref{I1I2}, which in this case are given explicitly by
\begin{equation}
\begin{split}
I_1^{(2)}=& \ Tr[{\omega^\bot}^2] \equiv Tr[({\hat{\omega}^\bot})^2]=\left( Tr\left[L^\dag \de_\mu L\tau_L^a\right]-\right.\left.Tr\left.[R^\dag\de_\mu R \tau_R^a \right] \right)^2,\\
I_2^{(2)}=& \ Tr[(\hat{\omega}^\parallel)^2]=\left(Tr\left[L^\dag \de_\mu L\tau_L^a\right]+\right.\left.Tr\left[R^\dag\de_\mu R \tau_R^a \right]-(iV_\mu^a) \right)^2.
\end{split}
\end{equation}
A further invariant,
\begin{equation}
\begin{split}
I_3^{(2)}=& \ Tr[{\hat{\omega}^\bot}_\mu \hat{\omega}^{\parallel \mu}] = \left( Tr\left[L^\dag \de_\mu L\tau_L^a\right]-Tr\left[R^\dag\de_\mu R \tau_R^a \right] \right)\\
& \cdot \left(Tr\left[L^\dag \de_\mu L\tau_L^a\right]+Tr\left[R^\dag\de_\mu R \tau_R^a \right]-(iV_\mu^a) \right),
\end{split}
\end{equation}
which mixes the orthogonal and the parallel components of the Maurer-Cartan form, can be written in principle; this additional term is not present in a theory with a generic $G \to H$ symmetry breaking, but only arises in very symmetric cases such as $SU(2) \otimes SU(2) \to SU(2)_D$. However, this term is $P_{LR}$-odd, so we are not going to consider it in the following.

Since order $\OO(p^3)$ invariants are ruled out by Lorentz invariance, the first corrections to the $O(p^2)$ Lagrangian will be of order $\OO(p^4)$. We will detail them in the following section.

\subsection{The $\OO(p^4)$ effective Lagrangian}
\label{p4}

The operators that we can use to build invariants are either of $\OO(p)$ ($\hat{\omega}_\mu^\bot$, $\hat{\omega}_\mu^\parallel$) or of $\OO(p^2)$ ($F_{\mu\nu}(V)$, $\hat{F}_{\mu\nu}(W)$, $\hat{F}_{\mu\nu}(B)$). As such, the $\OO(p^4)$ operators will come from combinations of traces of two, three and four operators. Six invariants can be build by combining two operators: the kinetic terms for the $V$, $W$ and $B$ fields and three other terms mixing $V$, $W$ and $B$. They are listed in table \ref{tabbi}.
\begin{table}[h]
\begin{center}
\begin{tabular}{|c|cl|}
\hline 
a) & $I_V^{(kin)}=$&$Tr[F_{\mu\nu}(V) F^{\mu\nu}(V)]$\\
\hline
b) & $I_W^{(kin)}=$&$Tr[F_{\mu\nu}(W) F^{\mu\nu}(W)]$\\
\hline
c) & $I_B^{(kin)}=$&$F_{\mu\nu}(B) F^{\mu\nu}(B)$\\
\hline
d) & $I_1^{EW} =$&$Tr[\hat{F}_{\mu\nu}(W) \hat{F}^{\mu\nu}(B)]$\\
\hline
e) & $I_2^{EW}=$&$Tr[F_{\mu\nu}(V) \hat{F}^{\mu\nu}(W)]$\\
\hline
f) & $I_3^{EW}=$&$Tr[F_{\mu\nu}(V) \hat{F}^{\mu\nu}(B)]$\\
\hline
\end{tabular}
\end{center}
\caption{Invariant operators built with two operators}
\label{tabbi}
\end{table}

Invariants containing three operators are also strongly constrained from the request of Lorentz invariance and from the trace properties of the $H^a$ generators (see appendix \ref{OpTraces}). Eight objects can be built; they are shown in table \ref{tabtril}.
\begin{table}[h]
\begin{center}
\begin{tabular}{|c|cl|}
\hline 
a) & $I_7^{(4)}=$&$Tr[F_{\mu\nu}(V)[\hat{\omega}^{\bot\mu},\hat{\omega}^{\bot\nu}]\,]$\\
\hline
b) & $I_8^{(4)}=$&$Tr[F_{\mu\nu}(V)[\hat{\omega}^{\parallel \mu},\hat{\omega}^{\parallel \nu}]\,]$\\
\hline
c) & $I_4^{EW}=$&$Tr[\hat{F}_{\mu\nu}(W)[\hat{\omega}^{\bot\mu},\hat{\omega}^{\bot\nu}]\,]$\\
\hline
d) & $I_5^{EW}=$&$Tr[\hat{F}_{\mu\nu}(W)[\hat{\omega}^{\parallel \mu},\hat{\omega}^{\parallel \nu}]\,]$\\
\hline
e) & $I_6^{EW}=$&$Tr[\hat{F}_{\mu\nu}(W) \, \hat{\omega}^{\parallel \mu}\hat{\omega}^{\bot \nu}]$\\
\hline
c) & $I_7^{EW}=$&$Tr[\hat{F}_{\mu\nu}(B)[\hat{\omega}^{\bot\mu},\hat{\omega}^{\bot\nu}]\,]$\\
\hline
d) & $I_8^{EW}=$&$Tr[\hat{F}_{\mu\nu}(B)[\hat{\omega}^{\parallel \mu},\hat{\omega}^{\parallel \nu}]\,]$\\
\hline
e) & $I_9^{EW}=$&$Tr[\hat{F}_{\mu\nu}(B) \, \hat{\omega}^{\parallel \mu}\hat{\omega}^{\bot \nu}]$\\
\hline
\end{tabular}
\end{center}
\caption{Invariants built with three operators}
\label{tabtril}
\end{table}
These invariants were, in part, already considered in \cite{Casalbuoni:1986vq}.

The terms with four operators can be expressed in two different basis, a ``bilinear'' one and a ``quadrilinear'' one (table \ref{basi4}); details of their derivation are shown in appendix \ref{OpTraces}.
\begin{table}[h]
\begin{center}
\begin{tabular}{|c|cl|}
\hline
c) & $\tilde{I}_1^{(4)}=$&$Tr\left[\hat{\omega}_\mu^\bot\hat{\omega}_\nu^\bot\right]Tr\left[\hat{\omega}^{\bot\mu}\hat{\omega}^{\bot\nu} \right]$ \\
\hline
d) & $\tilde{I}_2^{(4)}=$&$Tr\left[\hat{\omega}_\mu^\bot\hat{\omega}^{\bot\mu}\right]Tr\left[\hat{\omega}_\nu^\bot\hat{\omega}^{\bot\nu} \right]  $\\
\hline
e) & $\tilde{I}_3^{(4)}=$&$Tr\left[\hat{\omega}^\parallel_\mu\hat{\omega}^\parallel_\nu\right]Tr\left[\hat{\omega}^{\parallel\mu}\hat{\omega}^{\parallel\nu} \right]$  \\
\hline
f) & $\tilde{I}_4^{(4)}=$&$Tr\left[\hat{\omega}^\parallel_\mu\hat{\omega}^{\parallel\mu}\right]Tr\left[\hat{\omega}^\parallel_\nu\hat{\omega}^{\parallel\nu} \right]  $ \\
\hline
g) & $\tilde{I}_5^{(4)}=$&$Tr\left[\hat{\omega}^\bot_\mu\hat{\omega}^\bot_\nu\right]Tr\left[\hat{\omega}^{\parallel\mu}\hat{\omega}^{\parallel\nu} \right]$  \\
\hline
h) & $\tilde{I}_6^{(4)}=$&$Tr\left[\hat{\omega}^\bot_\mu\hat{\omega}^{\bot\mu}\right]Tr\left[\hat{\omega}^\parallel_\nu\hat{\omega}^{\parallel\nu}\right]$  \\
\hline
\end{tabular}\spa\spa\begin{tabular}{|c|cl|}
\hline
c) & $I_1^{(4)}=$&$Tr\left[\hat{\omega}_\mu^\bot\hat{\omega}^{\bot\mu}\hat{\omega}_\nu^\bot\hat{\omega}^{\bot\nu}\right]$ \\
\hline
d) & $I_2^{(4)}=$&$\frac12 Tr\left[\hat{\omega}_\mu^\bot\hat{\omega}_\nu^\bot \{\hat{\omega}^{\bot\mu},\hat{\omega}^{\bot\nu}\}\right] $\\
\hline
e) & $I_3^{(4)}=$&$Tr\left[\hat{\omega}^\parallel_\mu\hat{\omega}^{\parallel\mu}\hat{\omega}^\parallel_\nu\hat{\omega}^{\parallel\nu}\right]$ \\
\hline
f) & $I_4^{(4)}=$&$\frac12 Tr\left[\hat{\omega}^\parallel_\mu\hat{\omega}^\parallel_\nu\left\{\hat{\omega}^{\parallel\mu},\hat{\omega}^{\parallel\nu}\right\} \right] $\\
\hline
g) & $I_5^{(4)}=$&$Tr\left[\hat{\omega}^\bot_\mu\hat{\omega}^\bot_\nu\hat{\omega}^{\parallel\mu}\hat{\omega}^{\parallel\nu} \right]$\\
\hline
h) & $I_6^{(4)}=$&$Tr\left[\hat{\omega}^\bot_\mu\hat{\omega}^{\bot\mu}\hat{\omega}^\parallel_\nu\hat{\omega}^{\parallel\nu}\right]$\\
\hline
\end{tabular} \end{center}\caption{Bilinear (left) and quadrilinear (right) basis for invariants built with four operators}\label{basi4}
\end{table}

If we choose to use, for instance, the quadrilinear basis, the most general $\OO(p^4)$ Lagrangian is finally given by
\begin{equation}
\label{lagpquattro}
 \la^{(4)} = \la^{(2)} + \la^{(kin)}_{vec} + \sum_i \xi_i I^{EW}_i + \sum_i \Xi_i I^{(4)}_i,
\end{equation}
where
\begin{equation}
\la^{(2)}=-\frac{v^2}{2}\left[ Tr\left[\hat{\omega}_\mu^\bot \hat{\omega}^{\bot\mu} \right]+\alpha \, Tr[\hat{\omega}_\mu^\parallel \hat{\omega}^{\parallel\mu}]\right]
\end{equation}
is the $\OO(p^2)$ Lagrangian (note that $I_1^{(2)}$ coefficient is fixed by the requirement that the kinetic terms for the $\pi$ is canonically normalized; see sec. \ref{ungauge}), $\la^{(kin)}_{vec}$ contains the canonical kinetic terms of the vector fields $V$, $W$ and $B$, and, among the non-kinetic $\OO(p^4)$ contributions, we have kept the ones containing the electroweak fields field strengths separated. All free parameters, that is $\alpha$, the coupling constants $g$, $g'$ and $g''$ and the $\OO(p^4)$ coefficients $\xi_i$ and $\Xi_i$, are dimensionless.

\subsection{Unitary gauge}
\label{ungauge}

Even before switching on the electroweak interactions by gauging $SU(2) \otimes U(1)$, only three of the six scalar degrees of freedom that the theory contains are physical, while the others provide $V$ with its longitudinal component and its mass. For phenomenological applications, it is useful to see this in more detail. The two unitary matrices $L$ and $R$ can be written in exponential form
\begin{equation}
\dd L=\exp\left [i\frac{\pi_L^a \tau^a_L}{v_L}\right]\equiv e^{i\pi_L} \quad
\dd R=\exp\left [i\frac{\dd\pi_R^a \tau^a_R}{v_R}\right ]\equiv e^{i\pi_R}\label{2.32}
\end{equation} 
According to these definitions, the terms with up to two $\pi$ fields in $I_1^{(2)}$, $I_2^{(2)}$ can be rewritten as
\begin{align}
I_1^{(2)}& \supset -\frac1{4}\left( \frac{\de_\mu\pi_L^a\de^\mu\pi_L^a}{v_L^2}-\frac{\de_\mu\pi_L^a\de^\mu\pi_R^a}{v_Lv_R} \right) + (L\leftrightarrow R) \label{I1exp}\\
I_2^{(2)}& \supset \left( \frac{i\de_\mu\pi_L^a}{2v_L} - (iV_\mu^a) \right)^2 + (L\leftrightarrow R)
\end{align}
Note in particular that we have a mass term for the $V$ field along with a quadratic mixing term 
\begin{equation}
V^{\mu a} \left (\frac {\partial_\mu \pi^a_L}{v_L} + \frac {\partial_\mu \pi^a_R}{v_R}\right )
\end{equation}
which tell us that, as expected, that the Goldstone combinations providing mass for $V^a$ are $\frac{\pi^a_L}{v_L} + \frac{\pi^a_R}{v_R}$. 

The nonphysical degrees of freedom can be eliminated by a convenient gauge choice. Indeed, every $SU(2)_L\otimes SU(2)_R$ element can be rewritten as the product between an $SU(2)_L\otimes SU(2)_R/SU(2)_D$ element times an $SU(2)_D$ element, so that we have:
\[\left( \begin{array}{cc}
e^{i\pi_L} & 0\\
0 & e^{i\pi_R}
	\end{array} \right)=\left( \begin{array}{cc}
e^{i\pi} e^{i\sigma}& 0\\
0 & e^{-i\pi}e^{i\sigma}
					\end{array} \right)\]

If we do a gauge transformation with $h(x)=e^{-i\sigma}$  ($h(x)\in [SU(2)_V]_{loc}$), we cancel the $SU(2)_D$ contribution and we are left with just the coset contribution. This gauge choice can be realized simply by imposing the condition
\[\frac{\pi^a_L}{v_L}=-\frac{\pi^a_R}{v_R} \equiv \frac{\pi^a}{v}\]
where the $\pi^a$ now represent the physical fields, and for the sake of simplicity we will suppose that $v_L=v_R=v$, in such a way that every calculation made until now is automatically modified by substituting $\pi_L^a\fre\pi^a$ and $-\pi_R^a\fre\pi^a$. This eliminates by a matter of fact the quadratic mixing, along with three scalar degrees of freedom, and completely fixes the $[SU(2)_{V}]_{loc}$ gauge. In analogy with the SM, we will call this gauge choice the \emph{unitary gauge}.

Going to the unitary gauge drastically simplifies the Maurer-Cartan form expression:
\begin{equation}
\begin{array}{l}
\dd \hat{\omega}_\mu^\bot
= \de_\mu\pi+\frac1{3!}[\pi,[\pi,\de_\mu\pi]]+\dots\\
\dd \hat{\omega}_\mu^\parallel = -\frac1{2!}[\pi,\de_\mu\pi]-V_\mu +\dots\ \, ,
\end{array}
\label{omegariscritta}
\end{equation}
where $\pi\equiv \frac iv\pi^a H^a$. We will use these simplified expressions for $\hat{\omega}_\mu^\bot$, $\hat{\omega}_\mu^\parallel$ exclusively from now on.

\section{$\pi\pi\fre\pi\pi$ scattering and unitarity constraints}
\label{unitconstr}

A fundamental feature of an Higgsless model is that the new resonances help unitarize the scattering of longitudinally polarized $W$ and $Z$ bosons. In this section we will proceed to verify to which degree this effectively happens in the $SU(2)_L \otimes SU(2)_R$ with a single resonance. We will make use the $\OO(p^4)$ Lagrangian \eqref{lagpquattro} to study $\pi\pi\fre\pi\pi$ scattering; by the equivalence theorem \cite{Cornwall:1973tb,Vayonakis:1976vz,Chanowitz:1985hj} we are allowed to identify (up to corrections of order $O\left( \frac{M_{W,Z}^2}{E^2} \right)$ which are negligible near the UV cut-off of the model) the amplitudes involving Goldstone bosons $\pi$ with the corresponding amplitudes involving longitudinally polarized gauge bosons $W, \, Z$, obtained using the full $SU(2)_L\otimes U(1)_Y$ gauged Lagrangian. We then use the result to analyze in detail the unitarity properties of the theory, obtaining its cut-off energy as a function of the parameters.

By making use of eq. \eqref{omegariscritta}, and neglecting the $W_\mu^a$ and $B_\mu$ fields thanks to the equivalence theorem, the invariants $I_1^{(2)}$ and $I_2^{(2)}$ can be rewritten as
\begin{gather}
I_1^{(2)}=-\frac 1{v^2}{\de_\mu\pi^a\de^\mu\pi^a}+\frac{1}{3{v^4}}
\left( {\pi^a\pi^a\de_\mu\pi^b\de^\mu\pi^b-\pi^a\de_\mu\pi^a\pi^b\de^\mu\pi^b}\right) + \ldots\\
I_2^{(2)}= -\frac{1}{4{v^4}}\left( {\pi^a\pi^a\de_\mu\pi^b\de^\mu\pi^b-\pi^a\de_\mu\pi^a\pi^b\de^\mu\pi^b} \right)-V_\mu^aV^{\mu a} +\epsilon^{abc}\frac{\pi^a\de_\mu\pi^b V^{\mu c}}{v^2} + \ldots \, ,
\end{gather}
where the dots stand for terms containing more than four fields.
In a similar way we can rewrite in an explicit form the relevant $\OO(p^4)$ invariants. Limiting us, again, to terms with up to four fields:
\begin{gather}
\begin{split}
I_V^{(kin)}=&-2\left(\de_\mu V^c_\nu \de^\mu V^{\nu c}-\de_\mu V^c_\nu\de^\nu V^{\mu c} -2\de_\mu V^c_\nu\epsilon^{lmc}V^{\mu l} V^{\nu m}\right)\\
&-(V_{\mu}^aV^{\mu a}V_\nu^bV^{\nu b}-V_\mu^a V_\nu^a V^{\mu b}V^{\nu b}) + \ldots
\end{split}\\
I_1^{(4)}=\frac{1}{4v^4}\left( \de_\mu \pi^a\de^\mu\pi^a\de^\nu\pi^b\de_\nu\pi^b\right) + \ldots\\
\end{gather}
\begin{gather}
I_2^{(4)}=\frac{1}{2v^4}\left[\de_\mu \pi^a\de_\nu\pi^a\de^\mu\pi^b\de^\nu\pi^b\right] + \ldots\\
I_3^{(4)}=\frac{1}4V_\mu^a V^{\mu a}V_\nu^b V^{\nu b} + \ldots\\
I_4^{(4)}=\frac{1}2V_\mu^a V_\nu^aV^{\mu b} V^{\nu b} + \ldots\\
I_5^{(4)}=\frac{1}{4v^2}\left( \de_\mu\pi^a\de_\nu\pi^a V^{\mu b}V^{\nu b}-\de_\mu \pi^aV^{\mu a}\de_\nu\pi^b V^{\nu b}+\de_\mu\pi^a V^{\nu a}\de_\nu\pi^b V^{\mu b} \right) + \ldots\\
I_6^{(4)}=\frac{1}{4v^2}\de_\mu\pi^a\de^\mu\pi^a V_\nu ^bV^{\nu b} + \ldots\\
I_7^{(4)}=\frac{2\epsilon^{abc}}{v^2} \de_\mu V_\nu^a\de^\mu\pi^b\de^\nu\pi^c+\frac{1}{v^2} \left( V_\mu^a\de^\nu\pi^a V_\nu^b\de^\mu\pi^b -V_\mu^a\de^\mu \pi^aV_\nu^b\de^\nu\pi^b\right) + \ldots\\
\begin{split}
I_8^{(4)}=&\frac{1}{v^2}\left[ (\de_\mu V_\nu^a-\de_\nu V_\mu^a)\pi^a\de^\mu\pi^bV^{\nu b} \right)-\left( (\de_\mu V_\nu^a-\de_\nu V_\mu^a)\de^\mu\pi^aV^{\nu b}\pi^b\right]\\
&+2\epsilon^{abc} \de_\mu V_\nu^aV^{\mu b}V^{\nu c}+\left( V_\mu^aV^{\nu a}V^{\mu b}V_\nu^b-V_\mu^aV^{\mu a}V_\nu^b V^{\nu b} \right) + \ldots
\end{split}
\end{gather}
Using the above expressions we can derive Feynman propagators and interaction vertices with up to 4 fields containing the vector bosons $V$ and the Goldstone bosons $\pi$, up to $\OO(p^4)$ order. The full result is reported in appendix \ref{appB}. In the next section, we will make use of the Feynman rules to obtain the $\pi\pi \to \pi\pi$ scattering amplitude.

\subsection{Scattering matrix and partial waves}
\label{partialw}

The custodial $SU(2)_D$ symmetry implies that the amplitude for $\pi\pi\fre\pi\pi$ scattering has the following form:
\begin{equation}
-iA(\pi^i\pi^j\fre\pi^l\pi^m)=\calA(s,t,u)\delta^{ij}\delta^{lm}+\calA(t,s,u)\delta^{il}\delta^{jm}+\calA(u,t,s)\delta^{im}\delta^{jl};
\end{equation}
with a straightforward calculation, we obtain, at tree level:
\begin{equation}\label{Astu}
\begin{array}{rl}
\dd\mathcal A(s,t,u)=&\dd-iA(\pi^+\pi^-\fre\pi^0\pi^0)=\frac{s}{v^2}\left( 1-\frac{3}{4}\alpha \right)+\frac{\Xi_1}{2v^4}s^2+\frac{\Xi_2}{2v^4}\left( t^2+u^2 \right) \\
\dd+&\dd\frac{1}{4v^2}\alpha M_V^2\left[\frac{u-s}{t-M_V^2}+\frac{t-s}{u-M_V^2} \right]-\frac{{g''}^2\alpha\Xi_7}{v^2}\left[\frac{t(u-s)}{t-M_V^2}+\frac{u(t-s)}{u-M_V^2} \right]\\
+&\dd\frac{{g''}^2\Xi_7^2}{v^4}\left[\frac{t^2(u-s)}{t-M_V^2}+\frac{u^2(t-s)}{u-M_V^2} \right]
\end{array}
\end{equation} 
where $M_V=\sqrt{\alpha}g''v$.

We can now construct the scattering matrix. The relevant channels are $\pi^+\pi^-$, $\pi^\pm\pi^0$, $\frac{1}{\sqrt2}\pi^0\pi^0$ and $\pi^\pm\pi^\pm$; the matrix is given by ($ \calA_{s,t,u} \equiv \calA(s,t,u)$):
\begin{equation}
\mathcal M=\left( \begin{array}{c|cccc}
&\pi^+\pi^- & \frac{\dd\pi^0\pi^0}{\sqrt2}   &\pi^\pm\pi^0 & \pi^\pm\pi^\pm  \\
\hline
\pi^+\pi^-& \mathcal A_{s,t,u}+\mathcal A_{t,s,u} &  \frac{\dd\mathcal A_{s,t,u}}{\sqrt2}  &      &  /    \\
\frac{\dd\pi^0\pi^0}{\sqrt2}&  \frac{\dd\mathcal A_{s,t,u}}{\sqrt2}  & \frac{\dd\mathcal A_{s,t,u}+\mathcal A_{t,s,u}+\mathcal A_{u,t,s}}2   &  /   &   /   \\
\pi^\pm\pi^0& / &  /   &  \mathcal A_{t,s,u}   &  /    \\
\pi^\pm\pi^\pm& /  &  /   &  /   &   \mathcal A_{u,t,s}+\mathcal A_{t,s,u}   \\
\end{array} \right).
\end{equation} 

The scattering matrix can be diagonalized by switching to the total isospin basis. The eigenvalues, corresponding to the $I=0,1,2$ channels, are
\begin{equation}
\begin{array}{rl}
\dd T(0)=&\dd3 \calA(s,t,u)+\calA(t,s,u)+\calA(u,t,s)=\frac{2s}{v^2}\left( 1-\frac{3}{4}\alpha \right)\\
\spa\dd+&\dd\frac{\Xi_1}{2v^4}\left(3s^2+\frac{t^2+u^2}{2}\right)+\frac{\Xi_2}{2v^4}(s^2+t^2+u^2)\\
+&\dd\frac12\frac{\alpha M_V^2}{v^2}\left[\frac{u-s}{t-M_V^2}+\frac{t-s}{u-M_V^2} \right]-\frac{2\alpha \Xi_7{g''}^2}{v^2}\left[ \frac{t(u-s)}{t-M_V^2}+\frac{u(t-s)}{u-M_V^2} \right]\\
\dd+&\dd\frac{2\Xi_7^2 {g''}^2}{v^4}\left[\frac{t^2(u-s)}{t-M_V^2}+\frac{u^2(t-s)}{u-M_V^2} \right],
\end{array}
\end{equation} 
\begin{equation}
\begin{array}{rl}
\mspace{50mu} \dd T(1)=&\dd\mathcal A(t,s,u)-\mathcal A(u,t,s)=\frac{t-u}{v^2}\left( 1-\frac{3}{4}\alpha \right)+\left(\frac{\Xi_1}{2v^4}-\frac{\Xi_2}{2v^4}\right)(t^2-u^2)\\
\dd+&\dd\frac14\frac{\alpha M_V^2}{v^2}\left[2\frac{u-t}{s-M_V^2}+\frac{s-t}{u-M_V^2}-\frac{s-u}{t-M_V^2} \right]\\
\dd-&\dd\frac{\alpha \Xi_7{g''}^2}{v^2}\left[2\frac{s(u-t)}{s-M_V^2}+\frac{u(s-t)}{u-M_V^2}-\frac{t(s-u)}{t-M_V^2} \right]\\
\dd+&\dd\frac{\Xi_7^2 {g''}^2}{v^4}\left[2\frac{s^2(u-t)}{s-M_V^2}+\frac{u^2(s-t)}{u-M_V^2}-\frac{t^2(s-u)}{t-M_V^2} \right],
\end{array}
\end{equation} 
\begin{equation}
\begin{array}{rl}
\dd T(2)=&\dd\mathcal A(t,s,u)+\mathcal A(u,t,s)=\frac{t+u}{v^2}\left( 1-\frac{3}{4}\alpha \right)+\frac{\Xi_1}{2v^4}(t^2+u^2)\\
+&\dd\frac{\Xi_2}{4v^4}(2s^2+t^2+u^2)+\frac14\frac{\alpha M_V^2}{v^2}\left[\frac{s-u}{t-M_V^2} +\frac{s-t}{u-M_V^2}\right]\\
-&\dd\frac{\alpha \Xi_7{g''}^2}{v^2}\left[\frac{t(s-u)}{t-M_V^2}+\frac{u(s-t)}{u-M_V^2} \right]+\frac{\Xi_7^2 {g''}^2}{v^4}\left[ \frac{t^2(s-u)}{t-M_V^2}+\frac{u^2(s-t)}{u-M_V^2} \right].
\end{array}
\end{equation}

To analyze the model unitarity properties, we define the partial waves $a_l^I$:
\begin{equation}
a_l^I=\frac{1}{64\pi}\int_{-1}^1d(\cos\theta)P_l(\cos\theta)T(I).
\end{equation}
Note that this definition agrees with that of \cite{Barbieri:2008cc} but differs from that of \cite{Veltman:1976rt},
\begin{equation}
a_l(s)=\frac{1}{32\pi}\int_{-1}^1d(\cos\theta)P_l(\cos\theta)A(s,t,u);
\end{equation}
however, the different normalization of isospin eigenstates in \cite{Veltman:1976rt} brings a compensation of factors, by which we are able to compare without ambiguity  the amplitudes $A(s,t,u)$ (anyway unaffected by this different definition) and the partial waves derived from the scattering matrix \cite{Veltman:1976rt} or from the isospin amplitudes $T(I)$.

The leading partial waves at $\OO(p^4)$ are:
\begin{equation}
\begin{split}
a_0^0(s)&=\dd\frac1{64\pi}\left[ -\frac{2 M_V^4}{v^4} \left(\frac{1}{g''}-2\Xi_7g''\right)^2+\frac{4 M_V^4}{v^4} \left(\frac{1}{g''}-2\Xi_7g''\right)^2\log\left( 1+\frac{s}{M_V^2} \right)\right.\\
-&\frac{2 M_V^6}{v^4} \left(\frac{1}{g''}-2\Xi_7g''\right)^2\frac1s\log\left( 1+\frac s{M_V^2} \right)+\frac{4s}{v^2}\left[ 1-\frac{3M_V^2}{4v^2} \left(\frac{1}{g''}-2\Xi_7g''\right)^2 \right]\\
+&\left.\frac1{3v^4}\left( 11\Xi_1+14\Xi_2+16\Xi_7^2{g''}^2 \right)s^2\right],
\end{split}
\label{a00}
\end{equation}
\begin{equation}
\begin{split}
a_0^2(s)&=\frac1{64\pi}\left[ \frac{M_V^4}{v^4} \left(\frac{1}{g''}-2\Xi_7g''\right)^2-\frac{2 M_V^4}{v^4} \left(\frac{1}{g''}-2\Xi_7g''\right)^2\log\left( 1+\frac{s}{M_V^2} \right)\right.\\
+&\frac{M_V^6}{v^4} \left(\frac{1}{g''}-2\Xi_7g''\right)^2\frac1s\log\left( 1+\frac s{M_V^2} \right)-\frac{2s}{v^2}\left[ 1-\frac{3M_V^2}{4v^2} \left(\frac{1}{g''}-2\Xi_7g''\right)^2 \right]\\
+&\left.\frac2{3v^4}\left( \Xi_1+4\Xi_2-4\Xi_7^2{g''}^2 \right)s^2\right]
\end{split}
\label{a02}
\end{equation}
and
\begin{equation}
a_1^1=\frac1{64\pi}\frac{1}{M_V^2 -s} \left[ \sum_{n=-1}^3b_ns^n+\sum_{n=-2}^1c_ns^n\log\left( 1+\frac{s}{M_V^2} \right)\right],
\label{a11}
\end{equation} 
where
\begin{equation}
\begin{array}{ll}
b_{-1}=-2 \frac{M_V^8}{v^4} \left(\frac{1}{g''}-2\Xi_7g''\right)^2 & c_{-2}=2 \frac{M_V^{10}}{v^4} \left(\frac{1}{g''}-2\Xi_7g''\right)^2\\
b_0=-2 \frac{M_V^6}{v^4} \left(\frac{1}{g''}-2\Xi_7g''\right)^2 & c_{-1}=3 \frac{M_V^8}{v^4} \left(\frac{1}{g''}-2\Xi_7g''\right)^2\\
b_1=\frac{2M_V^2}{3v^2}\left( 1+\frac{23M_V^2}{4v^2} \left(\frac{1}{g''}-2\Xi_7g''\right)^2 \right) & c_{0}=-3 \frac{M_V^6}{v^4} \left(\frac{1}{g''}-2\Xi_7g''\right)^2\\
b_2=-\frac2{v^2}\left[ 1-\frac{3M_V^2}{4v^2} \left(\frac{1}{g''}-2\Xi_7g''\right)^2 \right] & c_{1}=-2\frac{M_V^4}{v^4} \left(\frac{1}{g''}-2\Xi_7g''\right)^2\\
b_3=\frac1{3v^4}\left( \Xi_1-\Xi_2+6\Xi_7^2{g''}^2 \right); &
\end{array}
\end{equation} 
the result at order $\OO(p)^2$ can be of course immediately recovered from eqs. \eqref{a00}, \eqref{a02} and \eqref{a11} by setting $\Xi_i \equiv 0$.

The most significant consequence of adding the $\OO(p^4)$ operators is the appearance in the partial waves of terms growing as $s^2$ which are absent at $\OO(p^2)$. Not surprisingly, these terms dominate at high energy, and to obtain a consistent picture, with the unitarity bound lying beyond the mass of the resonance $M_V$, we have to ask that they have either very small or vanishing coefficients. We will start the analysis by requiring that the coefficients are exactly zero; then, we will relax this assumption and try to obtain bounds on their possible values. 

\subsection{Analysis with vanishing $s^2$ terms}

In order to have vanishing coefficients for all $s^2$-proportional terms in eqs. \eqref{a00}, \eqref{a02} and \eqref{a11}, we have to ask that the $\Xi_i$ satisfy the following set of equations:  
\begin{align*}
\text{i) } & 11\Xi_1+14\Xi_2+16\Xi_7^2{g''}^2 =0;\\
\text{ii) } & \Xi_1+4\Xi_2-4\Xi_7^2{g''}^2=0;\\
\text{iii) } & \Xi_1-\Xi_2+6 \Xi_7^2 =0.
\end{align*}
It is easy to see that conditions i), ii) and iii) are simultaneously verified if and only if
\begin{equation}
\label{squadro}
\Xi_1=-2\Xi_2=-4\Xi_7^2{g''}^2.
\end{equation} 
Since $\Xi_1$ and $\Xi_2$ are present in the coefficients of the $s^2$ terms only, after imposing eq. \eqref{squadro}, the partial waves just depend on $M_V$, $g''$ and $\Xi_7$. A further simplification is possible by noting that $g''$ and $\Xi_7$ always appear in the combination $\left(\frac{1}{g''}-2\Xi_7g''\right)^2$; by defining an effective coupling:
\begin{equation}
\label{Geff}
\frac{1}{{G''}^2} = \left(\frac{1}{g''}-2\Xi_7g''\right)^2,
\end{equation}
we have that the partial waves depend on only 2 parameters, $M_V$ and $G''$. When $\Xi_i \equiv 0$ (the $\OO(p^2)$ limit), we have $G''=g''$. If the relation \eqref{squadro} holds, from the point of view of unitarity, the replacement $g'' \to G''$ is the \emph{only} effect of the $\OO(p^4)$ operators. The effect, though simple to describe, is however highly significant if $g'' \gg 1$ (as can be expected in an Higgsless theory): even a small value of $\Xi_7$ can lead to a value of $G''$ which is very different from $g''$, and consequently very different value of the cut-off energy $\Lambda$.

It is also interesting to note that the effective coupling $G''$ is directly related to the $V$ bosons decay width into $\pi\pi$;
using the expression for the trilinear vertex (see Appendix \ref{appB3}), it is straightforward to obtain:
\begin{equation}
\Gamma_V \equiv \Gamma (V \to \pi\pi)=
\frac{1}{48\pi}\left( \frac12\alpha g''-\frac{\Xi_7g''}{v^2}M_V^2 \right)^2M_V \equiv \frac{1}{192\pi} \frac{1}{{G''}^2} \frac{M_V^4}{v^4} M_V;
\end{equation}
that is, for a given value of $M_V$ the decay width is inversely proportional to ${G''}^2$.

\begin{figure}[h]
\begin{center}
\includegraphics[height=6.5cm]{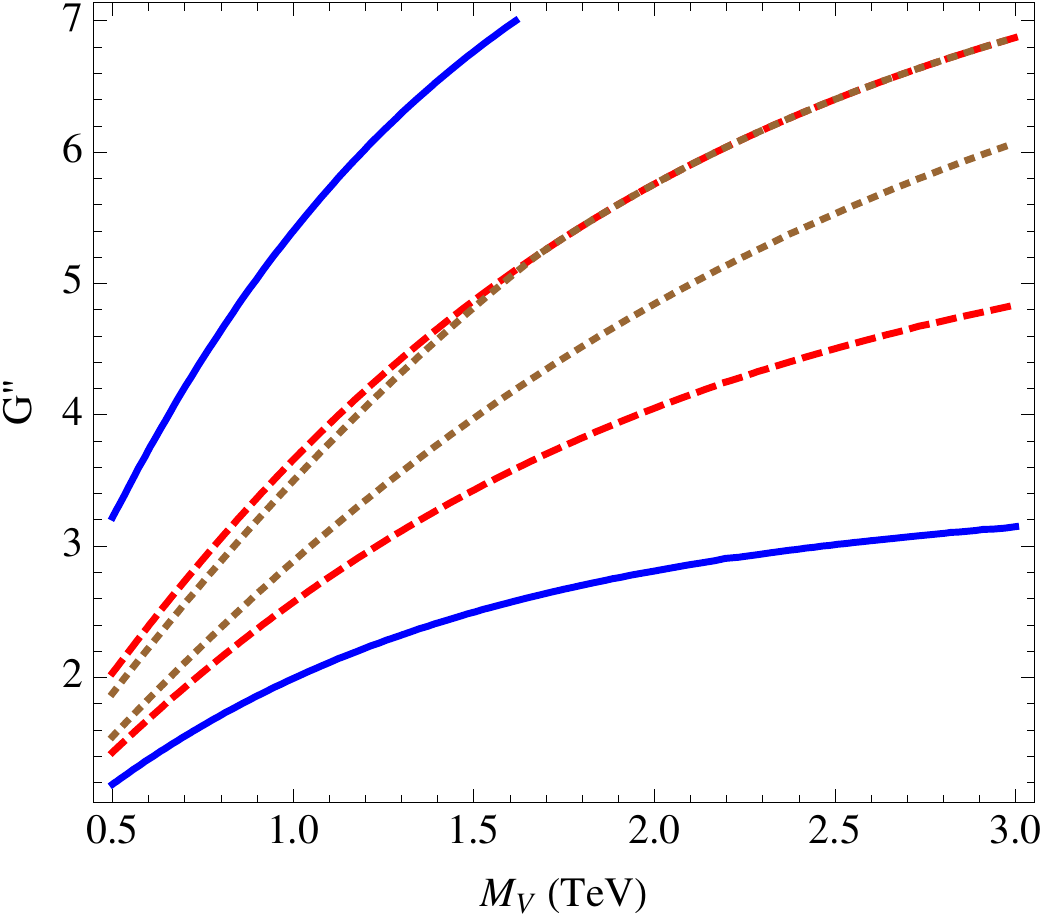}~~~~~~~
\includegraphics[height=6.5cm]{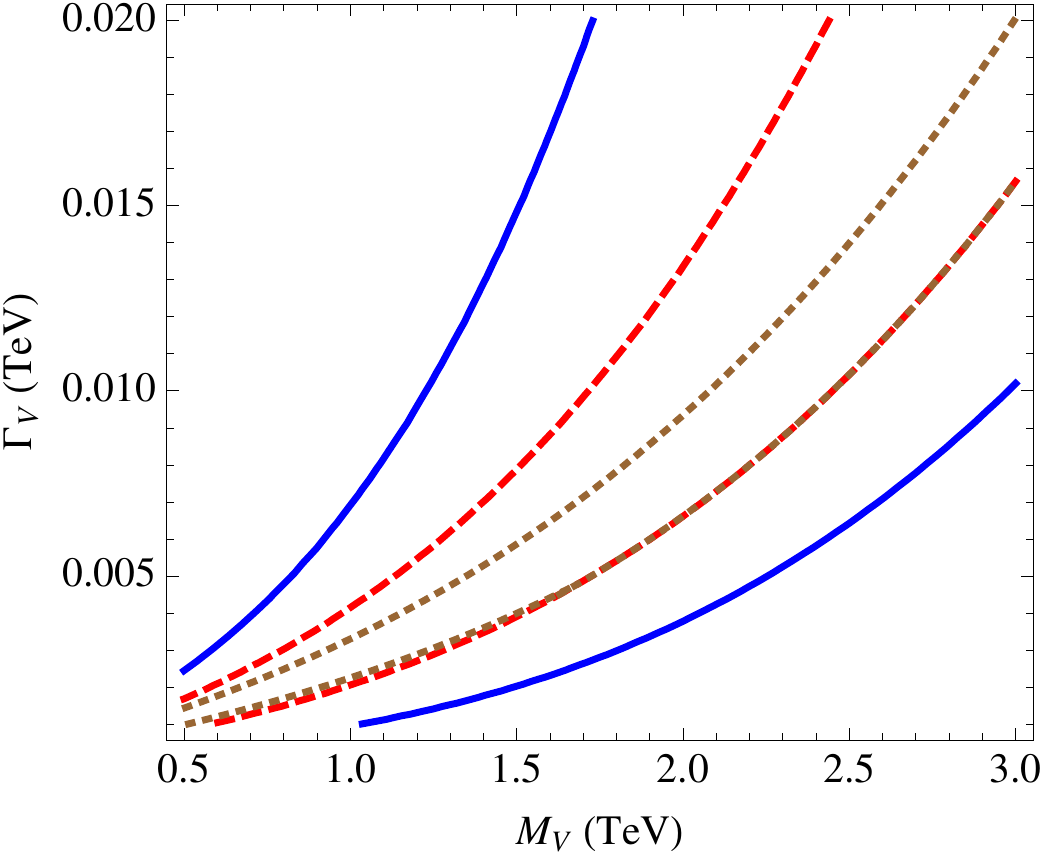}
\end{center}
\caption{\linespread{0.5}\small{Unitarity bounds in the parameter space $(M_V,G'')$ (on the left) and $(M_V,\Gamma_V)$ (on the right), using condition (\ref{squadro}) and requiring that $|a_l^I|<1$ up to 2 TeV (region between the solid blue lines), 3 TeV (between the dashed red lines) and 4 TeV (between the dotted brown lines).}\linespread{1.5}}
\label{fig:Cont1}
\end{figure}

\begin{figure}[h]
\begin{center}
\includegraphics[width=0.5\textwidth]{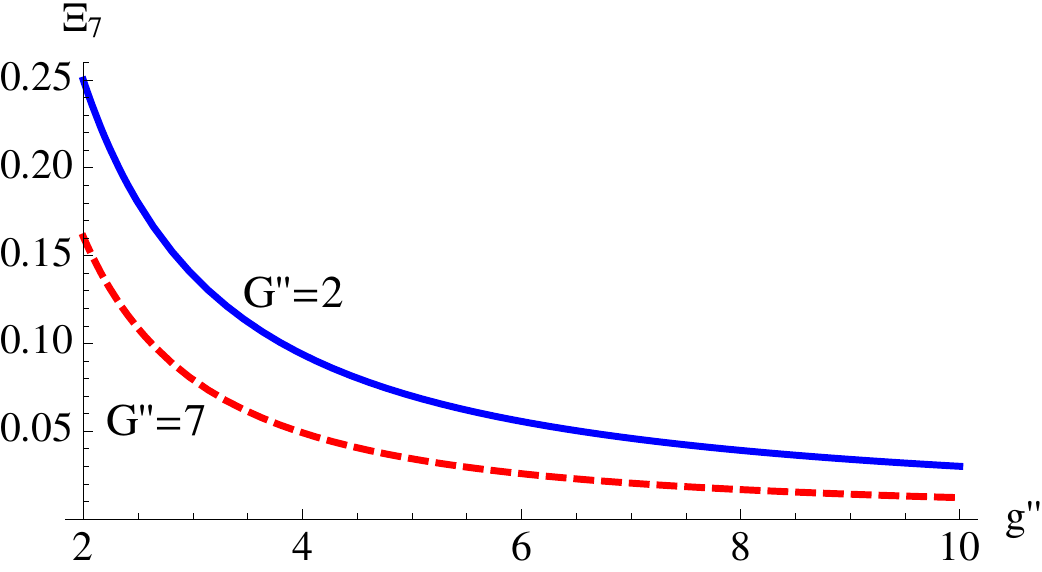}
\end{center}
\caption{\linespread{0.5}\small{$\Xi_7$ as a function of $g''$ in correspondence of two values of $G''$ in the region favoured by unitarity. The coefficient of the $\OO(p^4)$ operator has to be rather small, especially if $g'' > 4-5.$}\linespread{1.5}}
\label{fig:Xi7}
\end{figure}

In figure \ref{fig:Cont1} we show unitarity bounds in the $(M_V, G'')$ and $(M_V, \Gamma_V)$ planes. A sizable portion of parameter space is allowed even if we require that the unitarity violation is postponed beyond 3-4 TeV and that the new vector boson mass is $> 1$ TeV; $G''$ is constrained to be in general in the range $2-7$, with higher values preferred as the mass $M_V$ increases.
The implication from the point of view of the original operator coefficient, $\Xi_7$ is shown in fig. \ref{fig:Xi7}. As it can be seen, the values of $G''$ required to satisfy unitarity bounds imply that $\Xi_7$ is rather small, especially in correspondence of high values of $g''$.

\begin{figure}[t]
\begin{center}
\includegraphics[width=7.8cm]{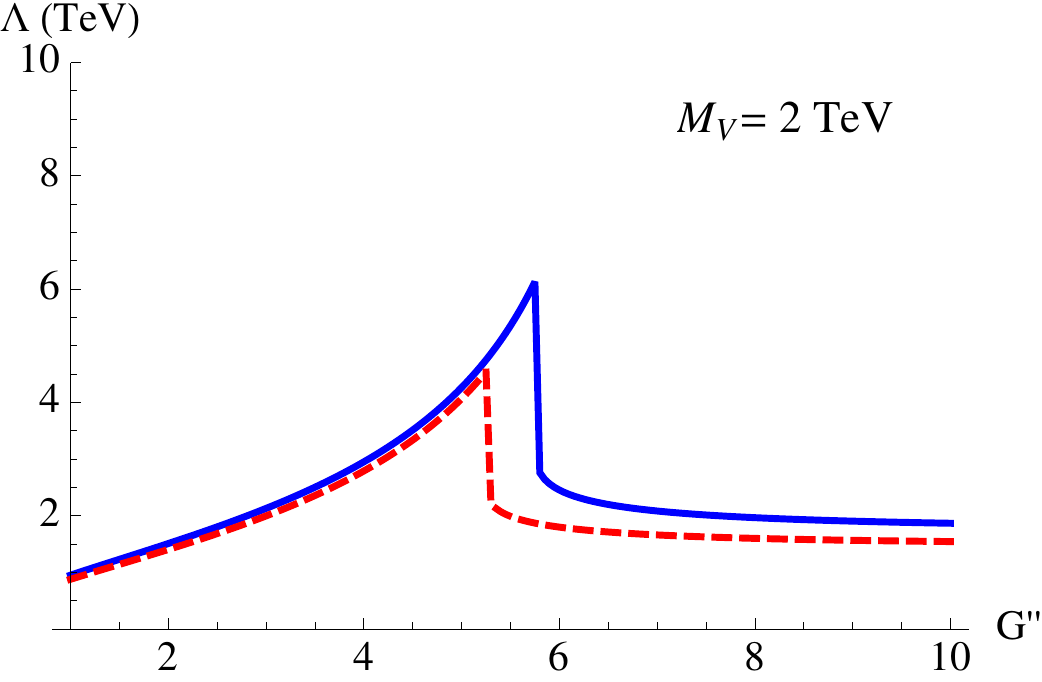}~~~~~~~
\includegraphics[width=7.8cm]{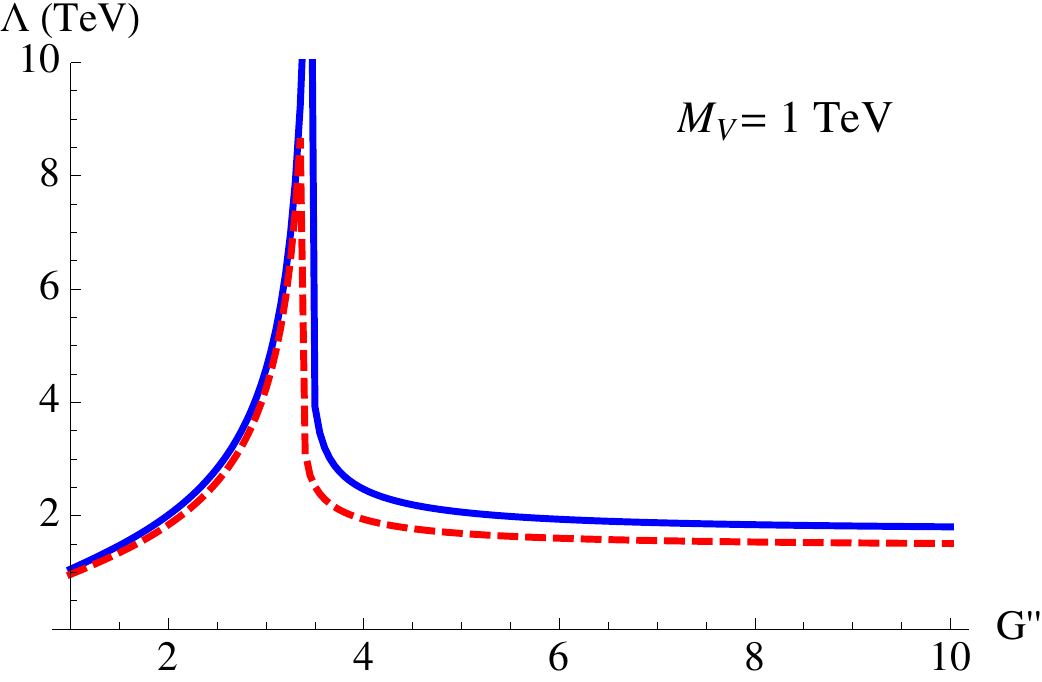}
\end{center}
\caption{\linespread{0.5}
\small{Unitarity bound, imposing condition \eqref{squadro}, obtained by requiring that $|a_l^I(s)| < 1$ (solid blue line) and $|a_l^I(s)| < 1/2$ (dotted red line) when $\sqrt{s}<\Lambda$, and fixing $M_V =1$ TeV (left) and $M_V=2$ TeV (right).}
\linespread{1.5}}
\label{fig:unilims}
\end{figure}
\begin{figure}[h]
\begin{center}
\includegraphics[width=7.8cm]{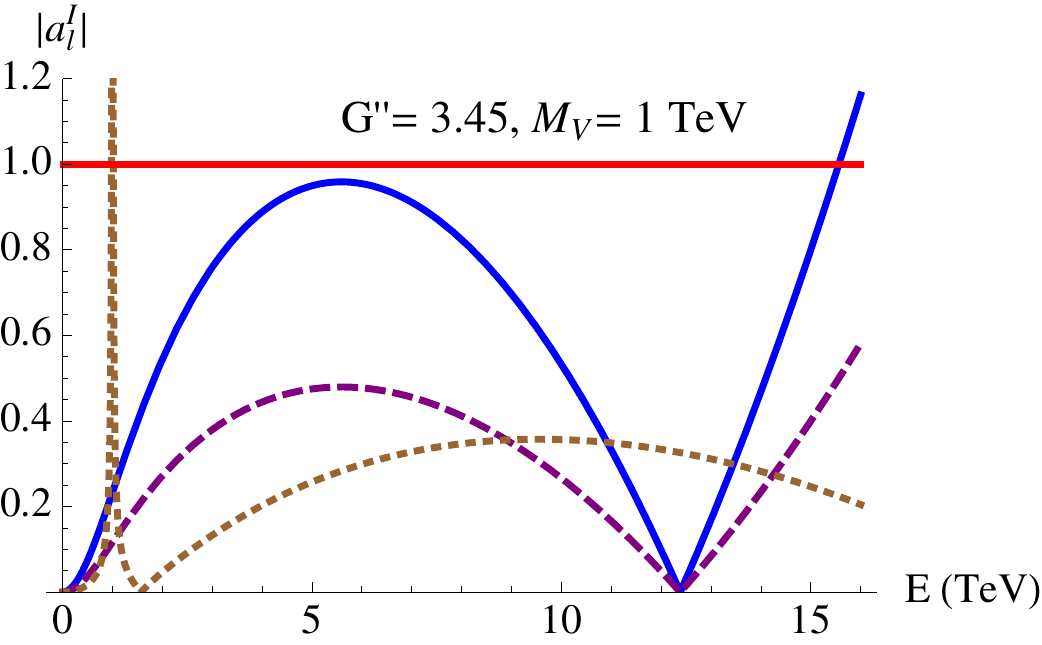}~~~~~~~
\includegraphics[width=7.8cm]{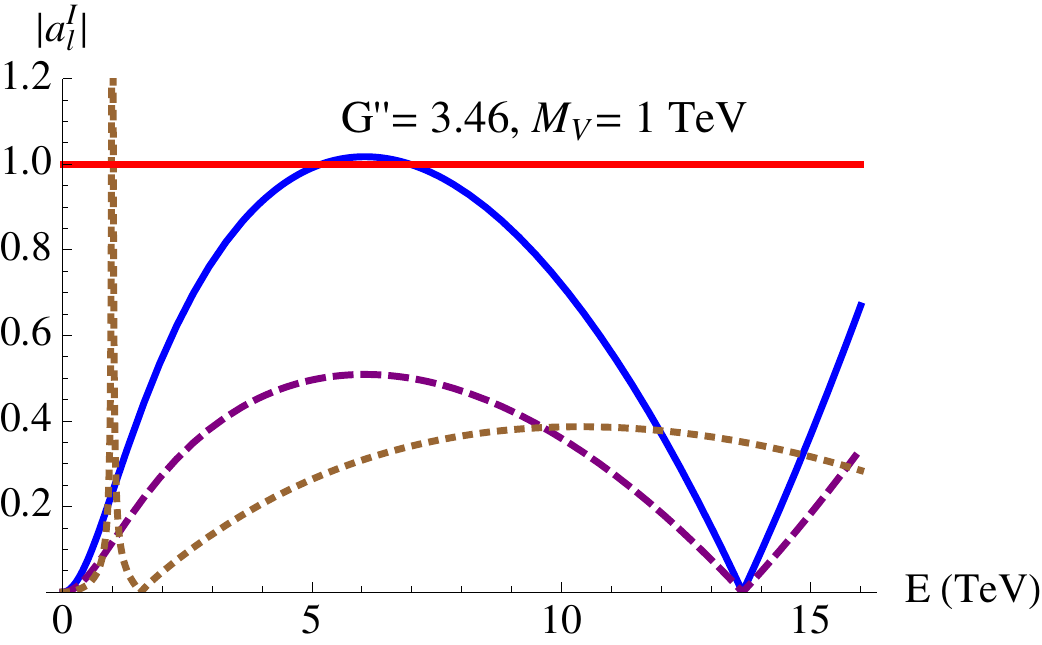}
\includegraphics[width=7.8cm]{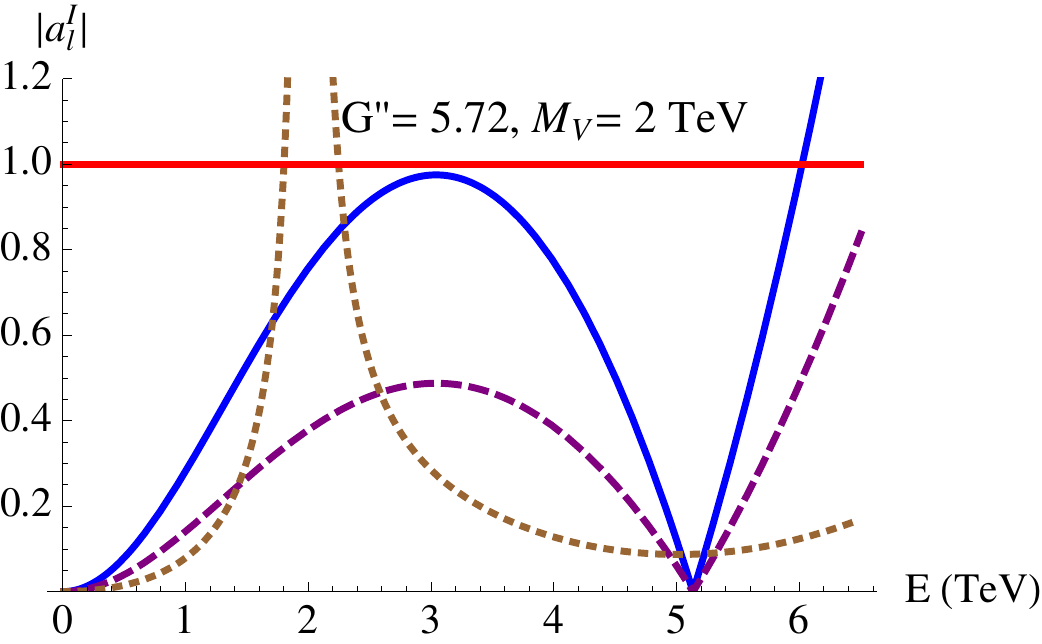}~~~~~~~
\includegraphics[width=7.8cm]{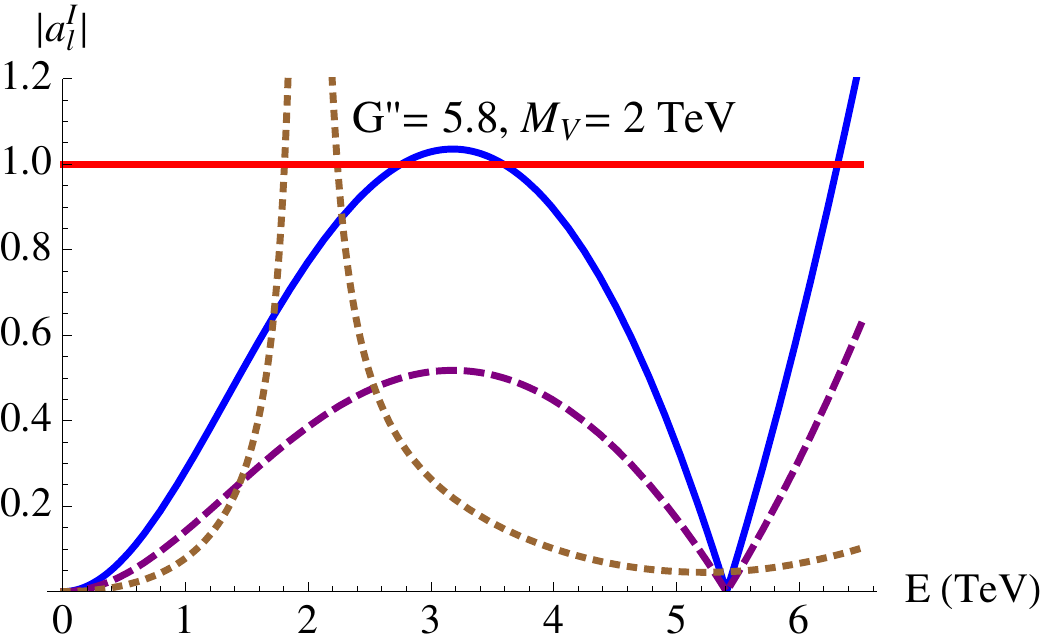}
\end{center}
\caption{\linespread{0.5}\small{Partial waves $a_0^0$ (blue, solid), $a_0^2$ (purple, dashed) and $a_0^1$ (grey, dotted), as a function of center-of-mass energy $E =\sqrt s$, using condition (\ref{squadro}). On the top: $M_V=1$ TeV and $G''=3.45$ (left) and $G''=3.46$ (right). On the bottom: $M_V=2$ TeV and $G''=5.72$ (left) and $G''=5.8$ (right). The values of $G''$ have been chosen in order to illustrate the behaviour near the discontinuity seen in fig. \ref{fig:unilims}.}\linespread{1.5}}
\label{fig:parwaves1}
\end{figure}

Having shown the limits on the parameter space, it is now interesting to calculate the actual unitarity limit corresponding to different fixed $M_V$ values, as a function of the effective coupling $G''$. We show the results in fig. \ref{fig:unilims}. As it can be seen, for $M_V$ in the range $1-2$ TeV, the limit is around 2 TeV in much of the parameter space, with a relatively narrow peak whose position depends on $M_V$. There is a discontinuity on the right side of the peak which can be better understood by looking at the energy dependence of the partial waves (fig. \ref{fig:parwaves1}) when $G''$ is chosen near the peak itself. It is apparent that the most stringent limit comes from the $a_0^0$ partial wave. In the region of larger cut-off, the coefficient of the dominant $s$-proportional term in $a_0^0$ is negative, and its contribution is partially compensated by the subdominant term $\propto \log(1+\frac{s}{M_V^2}))$ (which arises from integrating the vector propagator mass pole). As a result, the partial wave grows at low energy, then decreases (eventually becoming negative) as contribution of the $s$-term becomes more and more important. The first ramp goes higher at higher values of $G''$, and eventually it grows enough as to violate the unitarity bound. This effect is milder, but still present, if we choose impose $|a_l^I| < 1/2$ rather than $|a_l^I| <1$ as a condition.
By tuning $G''$ to reside just before the discontinuity, it is possible to push the cut-off to very high values, beyond 10 TeV when $M_V = 1$ TeV and up to 5-6 TeV when $M_V = 2$ TeV. However, as pointed out in \cite{Falkowski:2011ua}, at least at $\OO(p^2)$ these very high cut-off regions disappear when additional scattering channels, including the vector resonances $V$, are taken into account, even if sizable regions with a cut-off as high as $\sim 4$ TeV remain. This picture could change when the $\OO(p^4)$ operators are taken into account, as adding more scattering channels means introducing more free parameters in the calculation (remember that only three out of eight of the new invariants contribute to $\pi\pi \to \pi\pi$ scattering). The full analysis is beyond the scope of the present work; the main point we wish to stress here is that the contribution of the $\OO(p^4)$ operators, even with relatively small coefficients, can bring a quantitatively significant modification to the picture obtained at $\OO(p^2)$.

\subsection{Bounds on the $s^2$ terms}

\begin{figure}[t]
\begin{center}
\includegraphics[width=7.8cm]{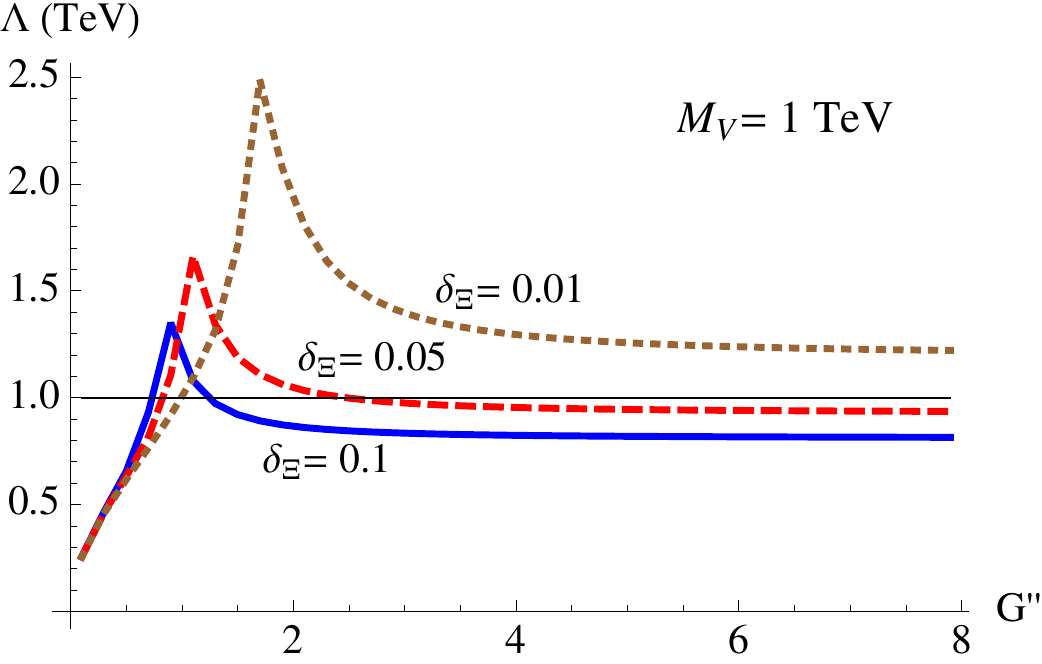}~~~~~~~
\includegraphics[width=7.8cm]{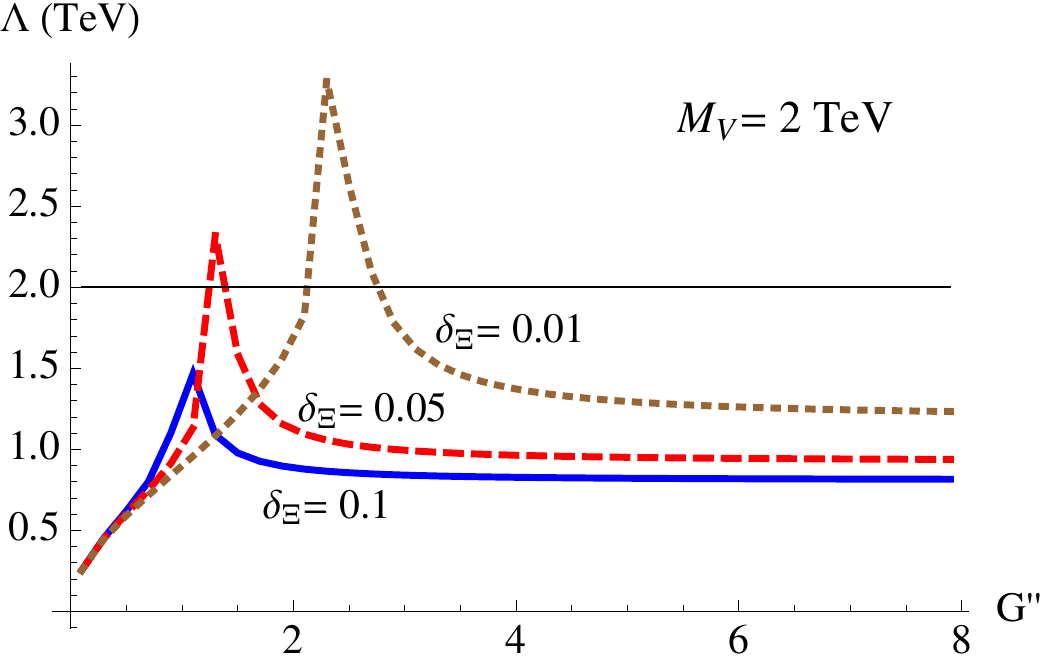}
\end{center}
\caption{\linespread{0.5}
\small{Unitarity bounds when the constraint \eqref{squadro} is not imposed, for different values of $\delta_\epsilon$. The straight black line represents the vector resonance mass. As it can be seen, most of the parameter space is ruled out unless $\delta_\epsilon$ is chosen to be very small.}
\linespread{1.5}}
\label{fig:s2plots}
\end{figure}

We will now evaluate the effect of the $s^2$-proportional terms and try to obtain bounds on their maximum values. Allowing nonzero $s^2$ terms introduces two additional variables, namely $\Xi_1$ and $\Xi_2$, in the study. An exhaustive search becomes more complicated and is also not very interesting for phenomenology. Instead, we will perform a simplified analysis using only the most stringent of the partial waves, $a_0^0$ (see fig. \ref{fig:parwaves1}), and expanding $\Xi_{1,2}$ around the values in eq. \eqref{squadro}:
\begin{equation}
\Xi_1 = -4 \Xi_7^2 {g''}^2 + \delta_\Xi, \quad \Xi_2 = 2 \Xi_7^2 {g''}^2 + \delta_\Xi;
\end{equation}
the expansion can be made in terms of a single parameter $\delta_\Xi$ since $\Xi_1$ and $\Xi_2$ appear in $a_0^0$ only in the $s^2$-proportional term; the partial waves becomes
\begin{equation}
a_0^0 = \frac{25}{192\pi} \frac{s^2}{v^4} \delta_\Xi + \OO(s).
\end{equation}
What we want to do is understand how large a value of $\delta_\Xi$ can be consistent with unitarity at least up to the scale of the resonance $V$. To do this, we plot the unitarity limit $\Lambda$ as a function of $G''$ for different values of $M_V$ and $\delta_\Xi$. The results are shown in fig. \ref{fig:s2plots}.
At $M_V=1$ TeV, we must have $\delta_\Xi < 0.1-0.05$ to have a limit $\Lambda>1$ TeV in a small narrow range of values for $G''$. At $\delta_\Xi = 0.01$, we have recovered most of the parameter space, but the unitarity behaviour is still much worse, in general, than the one corresponding to $\delta_\Xi = 0$. At $M_V=2$ TeV the situation is much worse: even when $\delta_\Xi = 0.01$, only a few values of $G''$ allow $\Lambda > 2$ TeV. The bottom line is that to have an unitarity behaviour which is significantly than in the SM without the Higgs, one needs condition \eqref{squadro} to hold at least in good approximation.

\section{Conclusions}
\label{conclusions}
We have studied a minimal Higgsless, including a single new vector resonance which is a triplet under the $SU(2)$ custodial symmetry of the SM, using an effective chiral lagrangian based on the $SU(2)\otimes SU(2) / SU(2)$ coset. We have written down the most general such lagrangian up to $\OO(p^4)$ and the associated set of Feynman rules, then used them to calculate the $\pi\pi \to \pi\pi$ scattering amplitude and study the unitarity properties of the model, obtaining an estimate for its cut-off energy scale $\Lambda$. We have particularly stressed the impact of the $\OO(p^4)$; in general, terms proportional to $s^2$ appear in the partial waves, and one has to ask that they at least approximately cancel each other in order to obtain a self-consistent picture (with the new resonance lying below the cut-off). Even when this cancellation is assumed, the remaining $\OO(p^4)$ contributions can have a significant effect, leading to a sizable modification of the estimate for the cut-off scale also in cases when the (dimensionless) coefficients of the corresponding operators are relatively small.

\appendix

\section{Operator traces}
\label{OpTraces}

The trace properties of the $H^a$, $X^a$ generators strongly constrain the invariants that can be built using the basic blocks containing either the scalars, $\omega_\mu^{\bot}$, $\hat{\omega}_\mu^\bot$ and $\hat{\omega}_\mu^{\parallel}$, or the vectors field strengths, $F_{\mu\nu}(V)$, $\hat{F}_{\mu\nu}(W)$, $\hat{F}_{\mu\nu}(B)$. We will need to consider the trace of products of two, three or four operators. The first one is directly given by the orthonormality property of the generators, eq. \eqref{norm}. Consider now the trace of the product of three operators; using the properties of the Pauli matrices, it is easy to show that
\begin{equation}
Tr[H^a H^b H^c] = Tr[X^a X^b H^c] = \frac{i}{4} \epsilon^{abc}; \quad Tr[X^a H^b H^c] = Tr[X^a X^b X^c] = 0.
\label{tr3}
\end{equation}
Finally, for the trace of four operators, the identity
\begin{equation}
Tr[G^aG^bG^cG^d]=Tr[G^aG^b]Tr[G^cG^d]-Tr[G^aG^c]Tr[G^bG^d]+Tr[G^aG^d]Tr[G^bG^c]
\label{tr4}
\end{equation}
holds, with $G^a = H^a, \, X^a$.

An interesting consequence of eqs. \eqref{norm}, \eqref{tr3} and \eqref{tr4}, is that \emph{any term with no more than four operators and containing an odd number of broken generators $X^a$ identically vanishes}. Furthermore, in terms with an \emph{even} number of broken generators $X^a$, the overall result does not change if we make the replacement $X^a \to H^a$. This means that any term containing an odd number of $\omega_\mu^\bot$ terms vanishes, while in ones that have an even number of $\omega_\mu^\bot$ terms,  we can safely replace $\omega_\mu^\bot \to \hat{\omega}_\mu^\bot$. In short, we can \emph{always} use $\hat{\omega}_\mu^\bot$ in place of $\omega_\mu^\bot$ to build the invariants.

Consider now the consequences of identity \eqref{tr4} on the building of terms with four operators. The useful building blocks to $\OO(p^4)$ are just $\hat{\omega}_\mu^\bot$ and $\hat{\omega}_\mu^\parallel$, since the vector field strengths all have dimension $p^2$. For every quadruple of operators $\hat{\omega}_1^\mu,\hat{\omega}_2^\nu,\hat{\omega}_3^\alpha,\hat{\omega}_4^\beta$ we should in principle consider the following invariants:
\begin{equation}
\begin{array}{l}
\dd Tr[\hat{\omega}_1^\mu\hat{\omega}_{2\mu}\hat{\omega}_3^\nu\hat{\omega}_{4\nu}],\,\,\text{non cyclic permutations}\\
\dd Tr[\hat{\omega}_1^\mu\hat{\omega}_2^\nu\hat{\omega}_{3\mu}\hat{\omega}_{4\nu}],\,\,\text{non cyclic permutations}\\
\dd Tr[\hat{\omega}_1^\mu\hat{\omega}_2^\nu\hat{\omega}_{3\nu}\hat{\omega}_{4\mu}],\,\,\text{non cyclic permutations}\\
\dd Tr[\hat{\omega}_1^\mu\hat{\omega}_{2\mu}]Tr[\hat{\omega}_3^\nu\hat{\omega}_{4\nu}],\,\,Tr[\hat{\omega}_1^\mu\hat{\omega}_{3\mu}]Tr[\hat{\omega}_2^\nu\hat{\omega}_{4\nu}],\,\,Tr[\hat{\omega}_1^\mu\hat{\omega}_{4\mu}]Tr[\hat{\omega}_2^\nu\hat{\omega}_{3\nu}]\\
\dd Tr[\hat{\omega}_1^\mu\hat{\omega}_2^\nu]Tr[\hat{\omega}_{3\mu}\hat{\omega}_{4\nu}],\,\,Tr[\hat{\omega}_1^\mu\hat{\omega}_3^\nu]Tr[\hat{\omega}_{2\mu}\hat{\omega}_{4\nu}],\,\,Tr[\hat{\omega}_1^\mu\hat{\omega}_4^\nu]Tr[\hat{\omega}_{2\mu}\hat{\omega}_{3\nu}]\\
\dd Tr[\hat{\omega}_1^\mu\hat{\omega}_2^\nu]Tr[\hat{\omega}_{3\nu}\hat{\omega}_{4\mu}],\,\,Tr[\hat{\omega}_1^\mu\hat{\omega}_3^\nu]Tr[\hat{\omega}_{2\nu}\hat{\omega}_{4\mu}],\,\,Tr[\hat{\omega}_1^\mu\hat{\omega}_4^\nu]Tr[\hat{\omega}_{2\nu}\hat{\omega}_{3\mu}]
\end{array}
\end{equation}
A first simplification is obtained by noticing that eq. \eqref{tr4} implies a symmetry under $a\leftrightarrow c$ or $b\leftrightarrow d$; using this, 3 of the 6 non cyclic permutation can be discarded:
\begin{gather}
Tr[\hat{\omega}_1\hat{\omega}_2\hat{\omega}_3\hat{\omega}_4]=\hat{\omega}_1^a\hat{\omega}_2^b\hat{\omega}_3^c\hat{\omega}_4^dTr[H^aH^bH^cH^d]=\hat{\omega}_1^a\hat{\omega}_2^b\hat{\omega}_3^c\hat{\omega}_4^dTr[H^aH^dH^cH^b]\equiv Tr[\hat{\omega}_1\hat{\omega}_3\hat{\omega}_4\hat{\omega}_2] \nonumber\\
Tr[\hat{\omega}_1\hat{\omega}_2\hat{\omega}_4\hat{\omega}_3]\equiv Tr[\hat{\omega}_1\hat{\omega}_3\hat{\omega}_4\hat{\omega}_2]\\
Tr[\hat{\omega}_1\hat{\omega}_3\hat{\omega}_2\hat{\omega}_4]\equiv Tr[\hat{\omega}_1\hat{\omega}_4\hat{\omega}_2\hat{\omega}_3] \nonumber
\end{gather}
Then, each block exists in four ``versions'', for example $\hat{\omega}^{\bot\mu}$, $\hat{\omega}^{\bot\nu}$, $\hat{\omega}^{\bot}_{\mu}$ and $\hat{\omega}^{\bot}_\nu$, and the possible combinations are all those which are allowed by $P_{LR}$ and Lorentz symmetries and by the trace properties of the generators $H^a$. We have the following possibilities:
\begin{table}[h]
\begin{center}
\begin{tabular}{|c|c|}
\hline 
$\left( \hat{\omega}^{\bot\mu},\hat{\omega}^{\bot\nu},\hat{\omega}^{\bot}_{\mu},\hat{\omega}^{\bot}_\nu \right)$ & $\left( \hat{\omega}^{\parallel\mu},\hat{\omega}^{\parallel\nu},\hat{\omega}^{\parallel}_{\mu},\hat{\omega}^{\parallel}_\nu \right)$\\
\hline
$\left( \hat{\omega}^{\bot\mu},\hat{\omega}^{\bot\nu},\hat{\omega}^{\parallel}_{\mu},\hat{\omega}^{\parallel}_\nu \right)$ & $\left( \hat{\omega}^{\bot\mu},\hat{\omega}^{\bot}_\mu,\hat{\omega}^{\parallel\nu},\hat{\omega}^{\parallel}_\nu \right)$ \\
\hline
\end{tabular}
\end{center}
\end{table}\\
\textit{A priori} we have in total $4\times(3+3+3+3+3+3)=54$ possible invariants; systematic use of eq. \eqref{tr4} allows us to lower this number down to 15 (tab. \ref{tabquadril}).
\begin{table}[h]
\begin{center}
\begin{tabular}{|c|c|c|c|}
\hline
1) & $Tr\left[\hat{\omega}_\mu^\bot\hat{\omega}_\nu^\bot\right]Tr\left[\hat{\omega}^{\bot\mu}\hat{\omega}^{\bot\nu} \right]$ & 2) & $Tr\left[{\hat{\omega}^\parallel}_\mu{\hat{\omega}^\parallel}_\nu\right]Tr\left[\hat{\omega}^{\parallel \mu}\hat{\omega}^{\parallel \nu} \right]$ \\
\hline
3) & $Tr\left[\hat{\omega}^\bot_\mu\hat{\omega}^{\bot\mu}\right]Tr\left[{\hat{\omega}^\parallel}_\nu\hat{\omega}^{\parallel \nu}\right]$ & 4)
& $Tr\left[\hat{\omega}^\bot_\mu\hat{\omega}^\bot_\nu\right]Tr\left[\hat{\omega}^{\parallel \mu}\hat{\omega}^{\parallel \nu} \right]$\\
\hline
5) & $Tr\left[\hat{\omega}_\mu^\bot\hat{\omega}^{\bot\mu}\right]Tr\left[\hat{\omega}_\nu^\bot\hat{\omega}^{\bot\nu} \right]  $ &6) & $Tr\left[{\hat{\omega}^\parallel}_\mu\hat{\omega}^{\parallel \mu}\right]Tr\left[{\hat{\omega}^\parallel}_\nu\hat{\omega}^{\parallel \nu} \right]  $\\
\hline
7) & $Tr\left[\hat{\omega}_\mu^\bot\hat{\omega}_\nu^\bot\hat{\omega}^{\bot\mu}\hat{\omega}^{\bot\nu} \right]$ & 8) & $Tr\left[{\hat{\omega}^\parallel}_\mu{\hat{\omega}^\parallel}_\nu\hat{\omega}^{\parallel \mu}\hat{\omega}^{\parallel \nu} \right]$ \\
\hline
9) & $Tr\left[\hat{\omega}_\mu^\bot\hat{\omega}^{\bot\mu}\hat{\omega}_\nu^\bot\hat{\omega}^{\bot\nu}\right]$ & 10) & $Tr\left[{\hat{\omega}^\parallel}_\mu\hat{\omega}^{\parallel \mu}{\hat{\omega}^\parallel}_\nu\hat{\omega}^{\parallel \nu}\right]$ \\
\hline
11) & $Tr\left[\hat{\omega}^\bot_\mu\hat{\omega}^\bot_\nu\hat{\omega}^{\parallel \mu}\hat{\omega}^{\parallel \nu} \right]$ & 
12) & $Tr\left[\hat{\omega}^\bot_\mu\hat{\omega}^\bot_\nu\hat{\omega}^{\parallel \nu}\hat{\omega}^{\parallel \mu} \right]$ \\
\hline
13) & $Tr\left[\hat{\omega}^\bot_\mu\hat{\omega}^{\bot\mu}{\hat{\omega}^\parallel}_\nu\hat{\omega}^{\parallel \nu}\right]$ & 
14)& $Tr\left[\hat{\omega}^\bot_\mu{\hat{\omega}^\parallel}_\nu\hat{\omega}^{\bot\mu}\hat{\omega}^{\parallel \nu}\right]$\\
\hline
15)&$Tr\left[\hat{\omega}^\bot_\mu{\hat{\omega}^\parallel}_\nu\hat{\omega}^{\bot\nu}\hat{\omega}^{\parallel \mu} \right]$ & &  \\
\hline
\end{tabular}
\end{center}
\caption{Invariant operators built with quadrilinear terms}\label{tabquadril}
\end{table}\\
The identity given in eq. \eqref{tr4} also allows us to rewrite the traces of four operators in terms of products of traces of two operators (tab. \ref{vincoli}). This provides us nine further constraints which lower down to six the number of independent invariants; this number cannot be further lowered, so we can choose as a basis either the couples of bilinear or the quadrilinear terms (tab. \ref{basi}).
\begin{table}[h]
 \begin{center}
\begin{tabular}{|c|}
\hline
$Tr\left[\hat{\omega}_\mu^\bot\hat{\omega}_\nu^\bot\hat{\omega}^{\bot\mu}\hat{\omega}^{\bot\nu} \right]=2Tr\left[\hat{\omega}_\mu^\bot\hat{\omega}_\nu^\bot\right]^2-Tr\left[\hat{\omega}_\mu^\bot\hat{\omega}^{\bot\mu}\right]^2$\\
\hline
$Tr\left[\hat{\omega}_\mu^\bot\hat{\omega}^{\bot\mu}\hat{\omega}_\nu^\bot\hat{\omega}^{\bot\nu}\right]=Tr\left[\hat{\omega}_\mu^\bot\hat{\omega}^{\bot\mu}\right]^2$\\
\hline
$Tr\left[{\hat{\omega}^\parallel}_\mu{\hat{\omega}^\parallel}_\nu\hat{\omega}^{\parallel \mu}\hat{\omega}^{\parallel \nu} \right]=2Tr\left[{\hat{\omega}^\parallel}_\mu{\hat{\omega}^\parallel}_\nu\right]^2-Tr\left[{\hat{\omega}^\parallel}_\mu\hat{\omega}^{\parallel \mu}\right]^2$\\
\hline
$Tr\left[{\hat{\omega}^\parallel}_\mu\hat{\omega}^{\parallel \mu}{\hat{\omega}^\parallel}_\nu\hat{\omega}^{\parallel \nu}\right]=Tr\left[{\hat{\omega}^\parallel}_\mu\hat{\omega}^{\parallel \mu}\right]^2$\\
\hline
$Tr\left[\hat{\omega}^\bot_\mu\hat{\omega}^\bot_\nu\hat{\omega}^{\parallel \mu}\hat{\omega}^{\parallel \nu} \right]=Tr\left[\hat{\omega}^\bot_\mu\hat{\omega}^\bot_\nu\right]Tr\left[\hat{\omega}^{\parallel \mu}\hat{\omega}^{\parallel \nu}\right]$\\
\hline
$Tr\left[\hat{\omega}^\bot_\mu\hat{\omega}^\bot_\nu\hat{\omega}^{\parallel \nu}\hat{\omega}^{\parallel \mu} \right]=Tr\left[\hat{\omega}^\bot_\mu\hat{\omega}^\bot_\nu\right]Tr\left[\hat{\omega}^{\parallel \mu}\hat{\omega}^{\parallel \nu}\right]$\\
\hline
$Tr\left[\hat{\omega}^\bot_\mu\hat{\omega}^{\bot\mu}{\hat{\omega}^\parallel}_\nu\hat{\omega}^{\parallel \nu}\right]=Tr\left[\hat{\omega}^\bot_\mu\hat{\omega}^{\bot\mu}\right]Tr\left[{\hat{\omega}^\parallel}_\nu\hat{\omega}^{\parallel \nu}\right]$\\
\hline
$Tr\left[\hat{\omega}^\bot_\mu{\hat{\omega}^\parallel}_\nu\hat{\omega}^{\bot\mu}\hat{\omega}^{\parallel \nu} \right]=-Tr\left[\hat{\omega}^\bot_\mu\hat{\omega}^{\bot\mu}\right]Tr\left[\hat{\omega}^{\parallel \nu}{\hat{\omega}^\parallel}_\nu\right]$\\
\hline
$Tr\left[\hat{\omega}^\bot_\mu{\hat{\omega}^\parallel}_\nu\hat{\omega}^{\bot\nu}\hat{\omega}^{\parallel \mu} \right]=-Tr\left[\hat{\omega}^\bot_\mu\hat{\omega}^{\bot\nu}\right]Tr\left[\hat{\omega}^{\parallel \mu}{\hat{\omega}^\parallel}_\nu\right]$\\
\hline
$Tr\left[\hat{\omega}^\bot_\mu\hat{\omega}^{\parallel \mu}\hat{\omega}^{\bot\nu}{\hat{\omega}^\parallel}_\nu \right]=-Tr\left[\hat{\omega}^\bot_\mu\hat{\omega}^{\bot\nu}\right]Tr\left[\hat{\omega}^{\parallel \mu}{\hat{\omega}^\parallel}_\nu\right]$\\
\hline
\end{tabular} 
\end{center}
\caption{Constraints coming from eq. \eqref{tr4}}
\label{vincoli}
\end{table}\\
\begin{table}[h]
\begin{center}
\begin{tabular}{|c|c|}
\hline
c) & $Tr\left[\hat{\omega}_\mu^\bot\hat{\omega}_\nu^\bot\right]Tr\left[\hat{\omega}^{\bot\mu}\hat{\omega}^{\bot\nu} \right]$ \\
\hline
d) & $Tr\left[\hat{\omega}_\mu^\bot\hat{\omega}^{\bot\mu}\right]Tr\left[\hat{\omega}_\nu^\bot\hat{\omega}^{\bot\nu} \right]  $\\
\hline
e) & $Tr\left[\hat{\omega}^\parallel_\mu\hat{\omega}^\parallel_\nu\right]Tr\left[\hat{\omega}^{\parallel\mu}\hat{\omega}^{\parallel\nu} \right]$  \\
\hline
f) & $Tr\left[\hat{\omega}^\parallel_\mu\hat{\omega}^{\parallel\mu}\right]Tr\left[\hat{\omega}^\parallel_\nu\hat{\omega}^{\parallel\nu} \right]  $ \\
\hline
g) & $Tr\left[\hat{\omega}^\bot_\mu\hat{\omega}^\bot_\nu\right]Tr\left[\hat{\omega}^{\parallel\mu}\hat{\omega}^{\parallel\nu} \right]$  \\
\hline
h) & $Tr\left[\hat{\omega}^\bot_\mu\hat{\omega}^{\bot\mu}\right]Tr\left[\hat{\omega}^\parallel_\nu\hat{\omega}^{\parallel\nu}\right]$  \\
\hline
\end{tabular}\spa\spa\begin{tabular}{|c|c|}
\hline
c) & $Tr\left[\hat{\omega}_\mu^\bot\hat{\omega}^{\bot\mu}\hat{\omega}_\nu^\bot\hat{\omega}^{\bot\nu}\right]$ \\
\hline
d) & $\frac12 Tr\left[\hat{\omega}_\mu^\bot\hat{\omega}_\nu^\bot \{\hat{\omega}^{\bot\mu},\hat{\omega}^{\bot\nu}\}\right] $\\
\hline
e) & $Tr\left[\hat{\omega}^\parallel_\mu\hat{\omega}^{\parallel\mu}\hat{\omega}^\parallel_\nu\hat{\omega}^{\parallel\nu}\right]$ \\
\hline
f) & $\frac12 Tr\left[\hat{\omega}^\parallel_\mu\hat{\omega}^\parallel_\nu\left\{\hat{\omega}^{\parallel\mu},\hat{\omega}^{\parallel\nu}\right\} \right] $\\
\hline
g) & $Tr\left[\hat{\omega}^\bot_\mu\hat{\omega}^\bot_\nu\hat{\omega}^{\parallel\mu}\hat{\omega}^{\parallel\nu} \right]$\\
\hline
h) & $Tr\left[\hat{\omega}^\bot_\mu\hat{\omega}^{\bot\mu}\hat{\omega}^\parallel_\nu\hat{\omega}^{\parallel\nu}\right]$\\
\hline
    \end{tabular} \end{center}\caption{Bilinear (left) and quadrilinear (right) basis}\label{basi}
\end{table}

\section{Feynman rules}
\label{appB}
From the Lagrangian  $O(p^4)$ given  in Eq.~(\ref{lagpquattro}) we can derive the Feynman rules (in the limit in which the equivalence theorem holds). In our notation
\begin{equation}
\epsilon^{abc}\fre -i\tilde\epsilon^{ijk}\spa (a,b,c=1,2,3;\,i,j,k=+,-,0)\spa \epsilon^{123}=1=\tilde\epsilon^{+-0}\end{equation} 
\subsection{Propagators}
\label{appB2}
\bi
\item $\pi^a$:
\[\]\begin{equation}
\parbox{20mm}{\begin{fmffile}{p1}
\begin{fmfgraph*}(40,1)
\fmfleft{l}
\fmfright{r}
\fmf{plain,label=$p$}{l,r}
\fmflabel{$a$}{l}
\fmflabel{$b$}{r}
\end{fmfgraph*}
\end{fmffile}}\spa\spa=\frac{i\delta^{ab}}{p^2+i\epsilon}                                                         \end{equation} 
\[\]
\item $V_\mu^a$ (in the unitary gauge):
\begin{equation}
\parbox{20mm}{\begin{fmffile}{p2}
\begin{fmfgraph*}(40,1)
\fmfleft{l}
\fmfright{r}
\fmf{photon,label=$q$}{l,r}
\fmflabel{$\mu,a$}{l}
\fmflabel{$\nu,b$}{r}

\end{fmfgraph*}
\end{fmffile}}\,\,\,\,=\frac{-i\delta_{ab}}{q^2-M_V^2+i\epsilon}\left( \eta_{\mu\nu}-\frac{q_\mu q_\nu}{M_V^2} \right)                                                                                                                      \end{equation} 

\ei
\subsection{Trilinear vertices}
\label{appB3}

\begin{equation}\boxed{
\la^{(2)}_{\pi\pi V}=\frac{i\alpha g''}{2}\tilde\epsilon^{ijk}\pi^i\de_\mu\pi^j V^{\mu k}}\spa\tilde\epsilon^{+-0}=1,\,i,j,k=+,-,0
\end{equation} 
\begin{equation}
\boxed{\la_{VII}^{\pi\pi V}=-2i\Xi_7g''\tilde\epsilon^{ijk}\left( \frac{1}{v} \right)^2\de^\mu\pi^i\de^\nu\pi^j\de_\mu V_\nu^k}                                                                                                                            \end{equation} 
\[
\put(40,20){(2+VII)}\parbox{30mm}{\begin{fmffile}{t1}
\begin{fmfgraph*}(60,60)
\fmfleft{l1,l2}
\fmfright{r}
\fmf{plain,label=$p_1$}{l1,o}
\fmf{plain,label=$p_2$,label.side=left}{l2,o}
\fmf{photon,label=$q$}{o,r}

\fmflabel{$i$}{l1}
\fmflabel{$j$}{l2}
\fmflabel{$k,\mu$}{r}

\end{fmfgraph*}
\end{fmffile}}\,\,\,\,=\frac{\delta i(\la^{(2)}+\la_{VII})}{\delta\pi^i(p_1)\delta\pi^j(p_2)\delta V^k_\mu(q)}\]\[
=i\epsilon^{ijk}\frac{g''}{2}\left [\left( \alpha -4\frac{\Xi_7}{v^2}(p_1\cdot p_2)\right)(p_2-p_1)^\nu+4\frac{\Xi_7}{v^2}(p_2^2p_1^\nu-p_1^2p_2^\nu)\right ]\]
\[\]
\[\boxed{\la_{kin}^{(3V)}+\la_{VIII}^{(3V)}=\left( g''+2\Xi_8{g''}^3 \right)\de_\mu V_\nu^aV^{\mu b}V^{\nu c}}\]

\[
\parbox{40mm}{\begin{fmffile}{t3}
\begin{fmfgraph*}(60,60)
\fmfleft{l1,l2}
\fmfright{r}
\fmf{photon,label=$q_1$}{l1,o}
\fmf{photon,label=$q_2$,label.side=left}{l2,o}
\fmf{photon,label=$q_3$}{o,r}

\fmflabel{$i,\mu$}{l1}
\fmflabel{$j,\nu$}{l2}
\fmflabel{$k,\lambda$}{r}

\end{fmfgraph*}
\end{fmffile}}\,\,\,\,=\frac{\delta i(\la^{kin}_V+\la_{VIII})}{\delta\pi^i(p_1)\delta\pi^j(p_2)\delta V^k(q)}\]
\[\]
\[
=i(g''+2{g''}^3\Xi_8)\tilde\epsilon^{ijk}\left( (q_2^\mu-q_3^\mu)\eta^{\lambda\nu}+(q_3^\nu-q_1^\nu)\eta^{\mu\lambda}+(q_1^\lambda-q_2^\lambda)\eta^{\mu\nu} \right)\]
\[\]
\subsection{Quadrilinear vertices}
\label{appB4}
\begin{equation}
\boxed{\begin{array}{rl}
\la^{(2)}_{4\pi}=&\dd -\frac{1}{2v^2}\left( \frac13-\frac{\alpha}{4} \right)\left[2\pi^+\pi^-\de_\mu\pi^+\de^\mu\pi^--\pi^+\pi^+\de_\mu\pi^-\de^\mu\pi^-\right.\\
-&\dd \pi^-\pi^-\de_\mu\pi^+\de^\mu\pi^+ + 2\pi^+\pi^-\de_\mu\pi^0\de^\mu\pi^0+2\pi^0\pi^0\de_\mu\pi^+\de^\mu\pi^-\\
-&\dd\left.2\pi^+\de_\mu\pi^-\pi^0\de^\mu\pi^0-2\pi^-\de_\mu\pi^+\pi^0\de^\mu\pi^0\right]
  \end{array}}              \end{equation} 
\[\]
\begin{equation}
\boxed{\la_I=\frac{\Xi_1}{v^4}\left( \de_\mu\pi^+\de^\mu\pi^-\de_\nu\pi^+\de^\nu\pi^-+\de_\mu\pi^0\de^\mu\pi^0 \de_\nu\pi^+\de^\nu\pi^-+\frac14\de_\mu\pi^0\de^\mu\pi^0 \de_\nu\pi^0\de^\nu\pi^0 \right)}                                                                                                                                                                                                            \end{equation}
\begin{equation}
\boxed{\begin{array}{rl}
\dd\la_{II}=&\dd\frac{\Xi_2}{v^4}\left( \de_\mu\pi^+\de^\mu\pi^+\de_\nu\pi^-\de^\nu\pi^-+\de_\mu\pi^+\de^\nu\pi^+\de_\nu\pi^-\de^\mu\pi^-\right.\\
+&\dd\left.2\de_\mu\pi^0\de^\mu\pi^+\de_\nu\pi^0 \de^\nu\pi^-+\frac12\de_\mu\pi^0\de^\mu\pi^0\de_\nu\pi^0 \de^\nu\pi^0 \right)       \end{array}}\end{equation} 
\[\]
\[
\put(-20,0){(2)}\parbox{20mm}{\begin{fmffile}{q1}
\begin{fmfgraph*}(60,60)
\fmfleft{l1,l2}
\fmfright{r1,r2}
\fmf{fermion, label=2,label.side=left}{l1,o1}
\fmf{fermion,label=1}{l2,o1}
\fmf{fermion,label=4,label.side=left}{o1,r1}
\fmf{fermion,label=3}{o1,r2}

\fmflabel{$a$}{l2}
\fmflabel{$b$}{l1}
\fmflabel{$c$}{r2}
\fmflabel{$d$}{r1}

\end{fmfgraph*}
\end{fmffile}}\spa=\frac{\delta i\la^{(2)}}{\delta\pi^a(p_1)\delta\pi^b(p_2)\delta\pi^c(p_3)\delta\pi^d(p_4)}\]
\[\]
\[=\frac{i}{v^2}\left( 1-\frac{3\alpha}{4} \right)\left( \delta^{ab}\delta^{cd}s+\delta^{ac}\delta^{bd}t+\delta^{ad}\delta^{bc}u \right)\]
where $\delta^{ab}=1$ only if  $ab=+-,-+,00$.
\[
\put(-20,0){(I)}\parbox{20mm}{\begin{fmffile}{q1}
\begin{fmfgraph*}(60,60)
\fmfleft{l1,l2}
\fmfright{r1,r2}
\fmf{fermion, label=2,label.side=left}{l1,o1}
\fmf{fermion,label=1}{l2,o1}
\fmf{fermion,label=4,label.side=left}{o1,r1}
\fmf{fermion,label=3}{o1,r2}

\fmflabel{$a$}{l2}
\fmflabel{$b$}{l1}
\fmflabel{$c$}{r2}
\fmflabel{$d$}{r1}

\end{fmfgraph*}
\end{fmffile}}\spa=\frac{\delta i\la_I}{\delta\pi^a(p_1)\delta\pi^b(p_2)\delta\pi^c(p_3)\delta\pi^d(p_4)}\]
\[\]
\[=i\frac{\Xi_1}{2v^4}\left( \delta^{ab}\delta^{cd}s^2+\delta^{ac}\delta^{bd}t^2+\delta^{ad}\delta^{bc}u^2 \right)\]
\[\]
\[
\put(-20,0){(II)}\parbox{20mm}{\begin{fmffile}{q1}
\begin{fmfgraph*}(60,60)
\fmfleft{l1,l2}
\fmfright{r1,r2}
\fmf{fermion, label=2,label.side=left}{l1,o1}
\fmf{fermion,label=1}{l2,o1}
\fmf{fermion,label=4,label.side=left}{o1,r1}
\fmf{fermion,label=3}{o1,r2}

\fmflabel{$a$}{l2}
\fmflabel{$b$}{l1}
\fmflabel{$c$}{r2}
\fmflabel{$d$}{r1}

\end{fmfgraph*}
\end{fmffile}}\spa=\frac{\delta i\la_I}{\delta\pi^a(p_1)\delta\pi^b(p_2)\delta\pi^c(p_3)\delta\pi^d(p_4)}\]
\[\]
\[=i\frac{\Xi_2}{2v^4}\left( \delta^{ab}\delta^{cd}(t^2+u^2)+\delta^{ac}\delta^{bd}(s^2+u^2)+\delta^{ad}\delta^{bc}(s^2+t^2) \right)\]
\[\]
\begin{equation}
\boxed{\begin{array}{rl}
\dd\la_{kin}^{4V}=&\dd-\frac{1}{2}{g''}^2\left( V_\mu^+V^{\mu-}V_\nu^+V^{\nu-}-V_\mu^+ V^{\mu+}V_\nu^-V^{\nu-}\right.\\
+&\dd\left.2\left( V_\mu^+V^{\mu-}V_\nu^0V^{\nu0}-V_\mu^+V^{\mu0}V_\nu^-V^{\nu0} \right) \right)       \end{array}}\end{equation}
\begin{equation}
\boxed{\begin{array}{rl}
\dd\la_{III}=&\dd\Xi_3{g''}^4\left(V_\mu^+V^{\mu-}V_\nu^+V^{\nu-}+V_\mu^+V^{\mu-}V_\nu^0V^{\nu0}\right.\\
+&\dd\left.\frac{1}{4}V_\mu^0V^{\mu0}V_\nu^0V^{\nu0}  \right)       \end{array}}     \end{equation} 
\begin{equation}
\boxed{\begin{array}{rl}
\dd\la_{IV}=&\dd\Xi_4{g''}^4\left(V_\mu^+V^{\mu-}V_\nu^+V^{\nu-}+V_\mu^+V^{\mu+}V_\nu^-V^{\nu-}+2V_\mu^+V^{\mu0}V_\nu^0V^{\nu-}\right.\\
+&\dd\left.\frac{1}{2}V_\mu^0V^{\mu0}V_\nu^0V^{\nu0}  \right)       \end{array}}\end{equation} 
\begin{equation}
\boxed{\begin{array}{rl}
\dd\la_{VIII}^{4V}=&\dd-\Xi_8{g''}^4\left(V_\mu^+V^{\mu-}V_\nu^+V^{\nu-}-V_\mu^+ V^{\mu+}V_\nu^-V^{\nu-}\right.\\
+&\dd\left.2\left( V_\mu^+V^{\mu-}V_\nu^0V^{\nu0}-V_\mu^+V^{\mu0}V_\nu^-V^{\nu0} \right)   \right)       \end{array}}                                                                                                                                                                                          \end{equation}
\[\]
\[
\put(-20,0){(kin)}\parbox{20mm}{\begin{fmffile}{q2}
\begin{fmfgraph*}(60,60)
\fmfleft{l1,l2}
\fmfright{r1,r2}
\fmf{photon, label=2,label.side=left}{l1,o1}
\fmf{photon,label=1}{l2,o1}
\fmf{photon,label=4,label.side=left}{o1,r1}
\fmf{photon,label=3}{o1,r2}

\fmflabel{$a,\mu$}{l2}
\fmflabel{$b,\nu$}{l1}
\fmflabel{$c,\alpha$}{r2}
\fmflabel{$d,\beta$}{r1}

\end{fmfgraph*}
\end{fmffile}}\spa=\frac{\delta i\la_{kin}^V}{\delta V_\mu^a(p_1)\delta V_\nu^b(p_2)\delta V_\alpha^c(p_3)\delta V_\beta^d(p_4)}\]
\[\]
\[=-i{g''}^2\left( \delta^{ab}\delta^{cd}\left( 2\eta_{\mu\nu}\eta_{\alpha\beta}-\eta_{\mu\alpha}\eta_{\nu\beta}-\eta_{\mu\beta}\eta_{\nu\alpha} \right)\right.\]
\[+\delta^{ac}\delta^{bd}\left( 2\eta_{\mu\alpha}\eta_{\nu\beta}-\eta_{\mu\nu}\eta_{\alpha\beta}-\eta_{\mu\beta}\eta_{\nu\alpha} \right)\]
\[\left.+\delta^{ad}\delta^{bc}\left( 2\eta_{\mu\beta}\eta_{\nu\alpha}-\eta_{\mu\alpha}\eta_{\nu\beta}-\eta_{\mu\nu}\eta_{\alpha\beta} \right) \right)\]
\[\]
\[
\put(-20,0){(III)}\parbox{20mm}{\begin{fmffile}{q2}
\begin{fmfgraph*}(60,60)
\fmfleft{l1,l2}
\fmfright{r1,r2}
\fmf{photon, label=2,label.side=left}{l1,o1}
\fmf{photon,label=1}{l2,o1}
\fmf{photon,label=4,label.side=left}{o1,r1}
\fmf{photon,label=3}{o1,r2}

\fmflabel{$a,\mu$}{l2}
\fmflabel{$b,\nu$}{l1}
\fmflabel{$c,\alpha$}{r2}
\fmflabel{$d,\beta$}{r1}

\end{fmfgraph*}
\end{fmffile}}\spa=\frac{\delta i\la_{III}}{\delta V_\mu^a(p_1)\delta V_\nu^b(p_2)\delta V_\alpha^c(p_3)\delta V_\beta^d(p_4)}\]
\[\]
\[=2i\Xi_3{g''}^4\left( \delta^{ab}\delta^{cd}\eta_{\mu\nu}\eta_{\alpha\beta}+\delta^{ac}\delta^{bd}\eta_{\mu\alpha}\eta_{\nu\beta}+ \delta^{ad}\delta^{bc}\eta_{\mu\beta}\eta_{\nu\alpha}\right)\]
\[\]
\[
\put(-20,0){(IV)}\parbox{20mm}{\begin{fmffile}{q2}
\begin{fmfgraph*}(60,60)
\fmfleft{l1,l2}
\fmfright{r1,r2}
\fmf{photon, label=2,label.side=left}{l1,o1}
\fmf{photon,label=1}{l2,o1}
\fmf{photon,label=4,label.side=left}{o1,r1}
\fmf{photon,label=3}{o1,r2}

\fmflabel{$a,\mu$}{l2}
\fmflabel{$b,\nu$}{l1}
\fmflabel{$c,\alpha$}{r2}
\fmflabel{$d,\beta$}{r1}

\end{fmfgraph*}
\end{fmffile}}\spa=\frac{\delta i\la_{III}}{\delta V_\mu^a(p_1)\delta V_\nu^b(p_2)\delta V_\alpha^c(p_3)\delta V_\beta^d(p_4)}\]
\[\]
\[\begin{array}{l}
=\dd2i\Xi_4{g''}^4\left( \delta^{ab}\delta^{cd}\left( \eta_{\mu\alpha}\eta_{\nu\beta} +\eta_{\mu\beta}\eta_{\nu\alpha}\right)+\delta^{ac}\delta^{bd}\left( \eta_{\mu\nu}\eta_{\alpha\beta} +\eta_{\mu\beta}\eta_{\nu\alpha}\right)\right.\\
\dd+\left. \delta^{ad}\delta^{bc}\left( \eta_{\mu\alpha}\eta_{\nu\beta} +\eta_{\mu\nu}\eta_{\beta\alpha}\right)\right)  \end{array}\]
\[\]
\[
\put(-30,0){(VIII)}\parbox{20mm}{\begin{fmffile}{q2}
\begin{fmfgraph*}(60,60)
\fmfleft{l1,l2}
\fmfright{r1,r2}
\fmf{photon, label=2,label.side=left}{l1,o1}
\fmf{photon,label=1}{l2,o1}
\fmf{photon,label=4,label.side=left}{o1,r1}
\fmf{photon,label=3}{o1,r2}

\fmflabel{$a,\mu$}{l2}
\fmflabel{$b,\nu$}{l1}
\fmflabel{$c,\alpha$}{r2}
\fmflabel{$d,\beta$}{r1}

\end{fmfgraph*}
\end{fmffile}}\spa=\frac{\delta i\la_{VIII}}{\delta V_\mu^a(p_1)\delta V_\nu^b(p_2)\delta V_\alpha^c(p_3)\delta V_\beta^d(p_4)}\]
\[\]
\[=-i\Xi_8{g''}^4\left( \delta^{ab}\delta^{cd}\left( 2\eta_{\mu\nu}\eta_{\alpha\beta}-\eta_{\mu\alpha}\eta_{\nu\beta}-\eta_{\mu\beta}\eta_{\nu\alpha} \right)\right.\]
\[+\delta^{ac}\delta^{bd}\left( 2\eta_{\mu\alpha}\eta_{\nu\beta}-\eta_{\mu\nu}\eta_{\alpha\beta}-\eta_{\mu\beta}\eta_{\nu\alpha} \right)\]
\[\left.+\delta^{ad}\delta^{bc}\left( 2\eta_{\mu\beta}\eta_{\nu\alpha}-\eta_{\mu\alpha}\eta_{\nu\beta}-\eta_{\mu\nu}\eta_{\alpha\beta} \right) \right)\]
\[\]
\begin{equation}
\boxed{\begin{array}{rl}
\la_{V}= &\dd-\Xi_5{g''}^2\left( \frac{i}{v} \right)^2\left( \de_\mu\pi^+ V^{\mu+}\de_\nu\pi^- V^{\mu-}+\frac14\de_\mu\pi^0V^{\mu0}\de_\nu\pi^0V^{\nu0}\right.\\
\dd+&\dd\frac12\de_\mu\pi^+V^{\mu0}\de_\nu\pi^-V^{\nu0}+\frac12\de_\mu\pi^0V^{\mu+}\de_\nu\pi^0 V^{\nu-}-\frac12\de_\mu\pi^+V^{\mu-}\de_\nu\pi^0V^{\nu0}\\
\dd-&\dd\left.\frac12\de_\mu\pi^-V^{\mu+}\de_\nu\pi^0V^{\nu0} +\frac12\de_\mu\pi^+V^{\mu0}\de_\nu\pi^0V^{\nu-}+\frac12\de_\mu\pi^-V^{\mu0}\de_\nu\pi^0V^{\mu+} \right)
  \end{array}}\end{equation} 
\[\]
\[
\put(-20,0){(V)}\parbox{20mm}{\begin{fmffile}{q3}
\begin{fmfgraph*}(60,60)
\fmfleft{l1,l2}
\fmfright{r1,r2}
\fmf{plain,label=2,label.side=left}{l1,o1}
\fmf{plain,label=1}{l2,o1}
\fmf{photon,label=4,label.side=left}{o1,r1}
\fmf{photon,label=3}{o1,r2}

\fmflabel{$a$}{l2}
\fmflabel{$b$}{l1}
\fmflabel{$c,\alpha$}{r2}
\fmflabel{$d,\beta$}{r1}

\end{fmfgraph*}
\end{fmffile}}\spa=\frac{\delta i\la_{V}}{\delta \pi^a(p_1)\delta \pi^b(p_2)\delta V_\alpha^c(p_3)\delta V_\beta^d(p_4)}\]
\[\]
\[\begin{array}{l}
\dd=-i\frac{\Xi_5{g''}^2}{2v^2}\left( \delta^{ab}\delta^{cd}\left( p_{1\alpha}p_{2\beta}+p_{2\alpha}p_{1\beta} \right)+\delta^{ac}\delta^{bd}\left( p_{2\alpha}p_{1\beta} -p_{1\alpha}p_{2\beta}\right)\right.\\
\dd\left.+\delta^{ad}\delta^{bc}\left( p_{1\alpha}p_{2\beta} -p_{2\alpha}p_{1\beta}\right) \right)  \end{array}\]
\[\]
\begin{equation}
\boxed{\begin{array}{rl}
\dd\la_{VI}=&\dd\frac{\Xi_6{g''}^2}{v^2}\left( \de_\mu\pi^+\de^\mu\pi^- V_\nu^+V^{\nu -}+\frac12\de_\mu\pi^+\de^\mu\pi^- V_\nu^0V^{\nu 0}+\frac12\de_\mu\pi^0\de^\mu\pi^0 V_\nu^+V^{\nu -}\right.\\
+&\dd\left.\frac14\de_\mu\pi^0\de^\mu\pi^0 V_\nu^0V^{\nu 0}\right)       \end{array}}                                  \end{equation}
\[\]
\[\]
\[
\put(-20,0){(VI)}\parbox{20mm}{\begin{fmffile}{q3}
\begin{fmfgraph*}(60,60)
\fmfleft{l1,l2}
\fmfright{r1,r2}
\fmf{plain,label=2,label.side=left}{l1,o1}
\fmf{plain,label=1}{l2,o1}
\fmf{photon,label=4,label.side=left}{o1,r1}
\fmf{photon,label=3}{o1,r2}

\fmflabel{$a$}{l2}
\fmflabel{$b$}{l1}
\fmflabel{$c,\alpha$}{r2}
\fmflabel{$d,\beta$}{r1}

\end{fmfgraph*}
\end{fmffile}}\spa=\frac{\delta i\la_{VI}}{\delta \pi^a(p_1)\delta \pi^b(p_2)\delta V_\alpha^c(p_3)\delta V_\beta^d(p_4)}=-i\frac{\Xi_6{g''}^2}{v^2}\delta^{ab}\delta^{cd}p_1\cdot p_2\eta_{\alpha\beta}\]
\[\]
\begin{equation}
\boxed{\begin{array}{rl}
\la_{VII}^{\pi\pi VV}=&\dd\frac{\Xi_7{g''}^2}{v^2}\left( V_\mu^+\de^\mu\pi^-V_\nu^+\de^\nu\pi^--V_\mu^-\de^\mu\pi^+V_\nu^-\de^\nu\pi^+\right.\\
+&\dd2V_\mu^+\de^\mu\pi^0V_\nu^0\de^\nu\pi^-+2V_\mu^-\de^\mu\pi^0V_\nu^0\de^\nu\pi^+-2V_\mu^+\de^\mu\pi^-V_\nu^0\de^\nu\pi^0\\
-&\dd\left.2V_\mu^-\de^\mu\pi^+V_\nu^0\de^\nu\pi^0\right)
\end{array} }                       \end{equation}
\[\]
\[\]
\[
\put(-20,0){(VII)}\parbox{20mm}{\begin{fmffile}{q3}
\begin{fmfgraph*}(60,60)
\fmfleft{l1,l2}
\fmfright{r1,r2}
\fmf{plain,label=2,label.side=left}{l1,o1}
\fmf{plain,label=1}{l2,o1}
\fmf{photon,label=4,label.side=left}{o1,r1}
\fmf{photon,label=3}{o1,r2}

\fmflabel{$a$}{l2}
\fmflabel{$b$}{l1}
\fmflabel{$c,\alpha$}{r2}
\fmflabel{$d,\beta$}{r1}

\end{fmfgraph*}
\end{fmffile}}\spa=\frac{\delta i\la_{VII}}{\delta \pi^a(p_1)\delta \pi^b(p_2)\delta V_\alpha^c(p_3)\delta V_\beta^d(p_4)}\]
\[\]
\[\]
\[=i\frac{2\Xi_7{g''}^2}{v^2}\left(\delta^{ac} \delta^{bd}\left(p_{1\alpha} p_{2\beta}-p_{2\alpha}p_{1\beta}  \right)+\delta^{ad} \delta^{bc}\left(p_{2\alpha} p_{1\beta}-p_{1\alpha}p_{2\beta}  \right) \right)\]
\[\]
\[\]
\begin{equation}
\boxed{\begin{array}{rl}
\la_{VIII}^{\pi\pi VV}=&\dd\Xi_8{g''}^2\left( \de_{[\mu}V^+_{\nu]}V^{\nu-}(\pi^-\de^\mu\pi^+-\pi^+\de^\mu\pi^-)\right.\\
\dd+&\dd\de_{[\mu}V^0_{\nu]}V^{\nu-}(\pi^0\de^\mu\pi^+-\pi^+\de^\mu\pi^0)+\de_{[\mu}V^-_{\nu]}V^{\nu+}(\pi^+\de^\mu\pi^--\pi^-\de^\mu\pi^+) \\
\dd+&\dd\de_{[\mu}V^0_{\nu]}V^{\nu+}(\pi^0\de^\mu\pi^--\pi^-\de^\mu\pi^0)+\de_{[\mu}V^+_{\nu]}V^{\nu0}(\pi^-\de^\mu\pi^0-\pi^0\de^\mu\pi^-)\\
+&\dd\left.\de_{[\mu}V^-_{\nu]}V^{\nu0}(\pi^+\de^\mu\pi^0-\pi^0\de^\mu\pi^+)\right)
  \end{array}}              \end{equation}
where
 $\de_{[\mu}V_{\nu]}\equiv\de_\mu V_\nu-\de_\nu V_\mu$.
\[\]
\[\]
\[
\put(-30,0){(VIII)}\parbox{20mm}{\begin{fmffile}{q3}
\begin{fmfgraph*}(60,60)
\fmfleft{l1,l2}
\fmfright{r1,r2}
\fmf{plain,label=2,label.side=left}{l1,o1}
\fmf{plain,label=1}{l2,o1}
\fmf{photon,label=4,label.side=left}{o1,r1}
\fmf{photon,label=3}{o1,r2}

\fmflabel{$a$}{l2}
\fmflabel{$b$}{l1}
\fmflabel{$c,\alpha$}{r2}
\fmflabel{$d,\beta$}{r1}

\end{fmfgraph*}
\end{fmffile}}\spa=\frac{\delta i\la_{VIII}}{\delta \pi^a(p_1)\delta \pi^b(p_2)\delta V_\alpha^c(p_3)\delta V_\beta^d(p_4)}\]
\[\]
\[\]
\[\begin{array}{l}
=\dd i\frac{\Xi_8{g''}^2}{v^2}\left(\delta^{ac} \delta^{bd}\left((p_4-p_3)\cdot(p_1-p_2)\eta_{\alpha\beta}-p_{4\alpha}(p_{1\beta}-p_{2\beta})-p_{3\beta}(p_{2\alpha}-p_{1\alpha})\right)\right.\\
\dd+\left.\delta^{ad} \delta^{bc}\left( -(p_4-p_3)\cdot(p_1-p_2)\eta_{\alpha\beta}+p_{4\alpha}(p_{1\beta}-p_{2\beta})+p_{3\beta}(p_{2\alpha}-p_{1\alpha}) \right) \right)  \end{array}\]
\[\]
\[\]
\[\]

\bibliography{biblio}

\begin{thebibliography}{10}

\bibitem{Weinberg:1978kz}
S.~Weinberg,
\newblock Physica {\bf A96}, 327 (1979).
%%CITATION = PHYSA,A96,327;%%

\bibitem{Gasser:1983yg}
J.~Gasser and H.~Leutwyler,
\newblock Ann. Phys. {\bf 158}, 142 (1984).
%%CITATION = APNYA,158,142;%%

\bibitem{Gasser:1984gg}
J.~Gasser and H.~Leutwyler,
\newblock Nucl. Phys. {\bf B250}, 465 (1985).
%%CITATION = NUPHA,B250,465;%%

\bibitem{Appelquist:1980vg}
T.~Appelquist and C.~W. Bernard,
\newblock Phys. Rev. {\bf D22}, 200 (1980).
%%CITATION = PHRVA,D22,200;%%

\bibitem{Longhitano:1980iz}
A.~C. Longhitano,
\newblock Phys. Rev. {\bf D22}, 1166 (1980).
%%CITATION = PHRVA,D22,1166;%%

\bibitem{Casalbuoni:1985kq}
R.~Casalbuoni, S.~De~Curtis, D.~Dominici, and R.~Gatto,
\newblock Phys. Lett. {\bf B155}, 95 (1985).
%%CITATION = PHLTA,B155,95;%%

\bibitem{Casalbuoni:1986vq}
R.~Casalbuoni, S.~De~Curtis, D.~Dominici, and R.~Gatto,
\newblock Nucl. Phys. {\bf B282}, 235 (1987).
%%CITATION = NUPHA,B282,235;%%

\bibitem{Casalbuoni:1988xm}
R.~Casalbuoni, S.~De~Curtis, D.~Dominici, F.~Feruglio, and R.~Gatto,
\newblock Int.J.Mod.Phys. {\bf A4}, 1065 (1989).

\bibitem{Lee:1977eg}
B.~W. Lee, C.~Quigg, and H.~B. Thacker,
\newblock Phys. Rev. {\bf D16}, 1519 (1977).
%%CITATION = PHRVA,D16,1519;%%

\bibitem{Antoniadis:1998ig}
I.~Antoniadis, N.~Arkani-Hamed, S.~Dimopoulos, and G.~R. Dvali,
\newblock Phys. Lett. {\bf B436}, 257 (1998), hep-ph/9804398.
%%CITATION = HEP-PH 9804398;%%

\bibitem{ArkaniHamed:1998rs}
N.~Arkani-Hamed, S.~Dimopoulos, and G.~R. Dvali,
\newblock Phys. Lett. {\bf B429}, 263 (1998), arXiv:hep-ph/9803315.
%%CITATION = HEP-PH/9803315;%%

\bibitem{Randall:1999ee}
L.~Randall and R.~Sundrum,
\newblock Phys. Rev. Lett. {\bf 83}, 3370 (1999), arXiv:hep-ph/9905221.
%%CITATION = HEP-PH/9905221;%%

\bibitem{Randall:1999vf}
L.~Randall and R.~Sundrum,
\newblock Phys. Rev. Lett. {\bf 83}, 4690 (1999), hep-th/9906064.
%%CITATION = HEP-TH 9906064;%%

\bibitem{Maldacena:1997re}
J.~M. Maldacena,
\newblock Adv. Theor. Math. Phys. {\bf 2}, 231 (1998), arXiv:hep-th/9711200.
%%CITATION = HEP-TH/9711200;%%

\bibitem{Gubser:1998bc}
S.~S. Gubser, I.~R. Klebanov, and A.~M. Polyakov,
\newblock Phys. Lett. {\bf B428}, 105 (1998), arXiv:hep-th/9802109.
%%CITATION = HEP-TH/9802109;%%

\bibitem{Witten:1998qj}
E.~Witten,
\newblock Adv. Theor. Math. Phys. {\bf 2}, 253 (1998), arXiv:hep-th/9802150.
%%CITATION = HEP-TH/9802150;%%

\bibitem{Csaki:2002gy}
C.~Csaki, J.~Erlich, and J.~Terning,
\newblock Phys. Rev. {\bf D66}, 064021 (2002), arXiv:hep-ph/0203034.
%%CITATION = HEP-PH/0203034;%%

\bibitem{Agashe:2003zs}
K.~Agashe, A.~Delgado, M.~J. May, and R.~Sundrum,
\newblock JHEP {\bf 08}, 050 (2003), hep-ph/0308036.
%%CITATION = HEP-PH 0308036;%%

\bibitem{Csaki:2003dt}
C.~Csaki, C.~Grojean, H.~Murayama, L.~Pilo, and J.~Terning,
\newblock Phys. Rev. {\bf D69}, 055006 (2004), arXiv:hep-ph/0305237.
%%CITATION = HEP-PH/0305237;%%

\bibitem{Csaki:2003sh}
C.~Csaki, C.~Grojean, J.~Hubisz, Y.~Shirman, and J.~Terning,
\newblock Phys. Rev. {\bf D70}, 015012 (2004), hep-ph/0310355.
%%CITATION = HEP-PH 0310355;%%

\bibitem{Nomura:2003du}
Y.~Nomura,
\newblock JHEP {\bf 11}, 050 (2003), hep-ph/0309189.
%%CITATION = HEP-PH 0309189;%%

\bibitem{SekharChivukula:2001hz}
R.~Sekhar~Chivukula, D.~A. Dicus, and H.-J. He,
\newblock Phys. Lett. {\bf B525}, 175 (2002), hep-ph/0111016.
%%CITATION = HEP-PH 0111016;%%

\bibitem{Ohl:2003dp}
T.~Ohl and C.~Schwinn,
\newblock Phys.Rev. {\bf D70}, 045019 (2004), arXiv:hep-ph/0312263.

\bibitem{Papucci:2004ip}
M.~Papucci,
\newblock (2004), hep-ph/0408058.
%%CITATION = HEP-PH 0408058;%%

\bibitem{Muck:2004br}
A.~Muck, L.~Nilse, A.~Pilaftsis, and R.~Ruckl,
\newblock Phys.Rev. {\bf D71}, 066004 (2005), arXiv:hep-ph/0411258.

\bibitem{Falkowski:2007iv}
A.~Falkowski, S.~Pokorski, and J.~P. Roberts,
\newblock JHEP {\bf 12}, 063 (2007), arXiv:0705.4653.
%%CITATION = 0705.4653;%%

\bibitem{ArkaniHamed:2001ca}
N.~Arkani-Hamed, A.~G. Cohen, and H.~Georgi,
\newblock Phys. Rev. Lett. {\bf 86}, 4757 (2001), hep-th/0104005.
%%CITATION = HEP-TH/0104005;%%

\bibitem{ArkaniHamed:2001nc}
N.~Arkani-Hamed, A.~G. Cohen, and H.~Georgi,
\newblock Phys. Lett. {\bf B513}, 232 (2001), arXiv:hep-ph/0105239.
%%CITATION = HEP-PH/0105239;%%

\bibitem{Hill:2000mu}
C.~T. Hill, S.~Pokorski, and J.~Wang,
\newblock Phys. Rev. {\bf D64}, 105005 (2001), hep-th/0104035.
%%CITATION = HEP-TH 0104035;%%

\bibitem{Cheng:2001vd}
H.-C. Cheng, C.~T. Hill, S.~Pokorski, and J.~Wang,
\newblock Phys. Rev. {\bf D64}, 065007 (2001), hep-th/0104179.
%%CITATION = HEP-TH 0104179;%%

\bibitem{Abe:2002rj}
H.~Abe, T.~Kobayashi, N.~Maru, and K.~Yoshioka,
\newblock Phys. Rev. {\bf D67}, 045019 (2003), hep-ph/0205344.
%%CITATION = HEP-PH/0205344;%%

\bibitem{Falkowski:2002cm}
A.~Falkowski and H.~D. Kim,
\newblock JHEP {\bf 08}, 052 (2002), hep-ph/0208058.
%%CITATION = HEP-PH 0208058;%%

\bibitem{Randall:2002qr}
L.~Randall, Y.~Shadmi, and N.~Weiner,
\newblock JHEP {\bf 01}, 055 (2003), hep-th/0208120.
%%CITATION = HEP-TH 0208120;%%

\bibitem{Son:2003et}
D.~T. Son and M.~A. Stephanov,
\newblock Phys. Rev. {\bf D69}, 065020 (2004), hep-ph/0304182.
%%CITATION = HEP-PH 0304182;%%

\bibitem{deBlas:2006fz}
J.~de~Blas, A.~Falkowski, M.~Perez-Victoria, and S.~Pokorski,
\newblock JHEP {\bf 08}, 061 (2006), hep-th/0605150.
%%CITATION = HEP-TH/0605150;%%

\bibitem{SekharChivukula:2006cg}
R.~Sekhar~Chivukula {\em et~al.},
\newblock Phys. Rev. {\bf D74}, 075011 (2006), arXiv:hep-ph/0607124.
%%CITATION = HEP-PH/0607124;%%

\bibitem{Foadi:2007ue}
R.~Foadi, M.~T. Frandsen, T.~A. Ryttov, and F.~Sannino,
\newblock Phys. Rev. {\bf D76}, 055005 (2007), arXiv:0706.1696.
%%CITATION = 0706.1696;%%

\bibitem{Barbieri:2008cc}
R.~Barbieri, G.~Isidori, V.~S. Rychkov, and E.~Trincherini,
\newblock (2008), arXiv:0806.1624.
%%CITATION = 0806.1624;%%

\bibitem{Barbieri:2009tx}
R.~Barbieri, A.~E. Carcamo~Hernandez, G.~Corcella, R.~Torre, and
  E.~Trincherini,
\newblock JHEP {\bf 03}, 068 (2010), arXiv:0911.1942.
%%CITATION = 0911.1942;%%

\bibitem{Accomando:2008jh}
E.~Accomando, S.~De~Curtis, D.~Dominici, and L.~Fedeli,
\newblock Phys.Rev. {\bf D79}, 055020 (2009), arXiv:0807.5051.

\bibitem{Falkowski:2011ua}
A.~Falkowski, C.~Grojean, A.~Kaminska, S.~Pokorski, and A.~Weiler,
\newblock (2011), arXiv:1108.1183,
\newblock * Temporary entry *.

\bibitem{Dawson:2007yk}
S.~Dawson and C.~B. Jackson,
\newblock Phys. Rev. {\bf D76}, 015014 (2007), arXiv:hep-ph/0703299.
%%CITATION = HEP-PH/0703299;%%

\bibitem{Abe:2008hb}
T.~Abe, S.~Matsuzaki, and M.~Tanabashi,
\newblock (2008), arXiv:0807.2298.
%%CITATION = 0807.2298;%%

\bibitem{Tanabashi:1993sr}
M.~Tanabashi,
\newblock Phys.Lett. {\bf B316}, 534 (1993), arXiv:hep-ph/9306237.

\bibitem{Contino:2011np}
R.~Contino, D.~Marzocca, D.~Pappadopulo, and R.~Rattazzi,
\newblock (2011), arXiv:1109.1570.

\bibitem{Balachandran:1978pk}
A.~P. Balachandran, A.~Stern, and C.~G. Trahern,
\newblock Phys. Rev. {\bf D19}, 2416 (1979).
%%CITATION = PHRVA,D19,2416;%%

\bibitem{Bando:1984ej}
M.~Bando, T.~Kugo, S.~Uehara, K.~Yamawaki, and T.~Yanagida,
\newblock Phys. Rev. Lett. {\bf 54}, 1215 (1985).
%%CITATION = PRLTA,54,1215;%%

\bibitem{Bando:1984pw}
M.~Bando, T.~Kugo, and K.~Yamawaki,
\newblock Prog. Theor. Phys. {\bf 73}, 1541 (1985).
%%CITATION = PTPKA,73,1541;%%

\bibitem{Bando:1985rf}
M.~Bando, T.~Kugo, and K.~Yamawaki,
\newblock Nucl. Phys. {\bf B259}, 493 (1985).
%%CITATION = NUPHA,B259,493;%%

\bibitem{Bando:1987br}
M.~Bando, T.~Kugo, and K.~Yamawaki,
\newblock Phys. Rept. {\bf 164}, 217 (1988).
%%CITATION = PRPLC,164,217;%%

\bibitem{Coleman:1969sm}
S.~R. Coleman, J.~Wess, and B.~Zumino,
\newblock Phys.Rev. {\bf 177}, 2239 (1969).

\bibitem{Callan:1969sn}
J.~Callan, Curtis~G., S.~R. Coleman, J.~Wess, and B.~Zumino,
\newblock Phys.Rev. {\bf 177}, 2247 (1969).

\bibitem{Ecker:1989yg}
G.~Ecker, J.~Gasser, H.~Leutwyler, A.~Pich, and E.~de~Rafael,
\newblock Phys.Lett. {\bf B223}, 425 (1989).

\bibitem{Ecker:1988te}
G.~Ecker, J.~Gasser, A.~Pich, and E.~de~Rafael,
\newblock Nucl.Phys. {\bf B321}, 311 (1989).

\bibitem{Pallante:1992qe}
E.~Pallante and R.~Petronzio,
\newblock Nucl.Phys. {\bf B396}, 205 (1993).

\bibitem{Bijnens:1995ii}
J.~Bijnens and E.~Pallante,
\newblock Mod.Phys.Lett. {\bf A11}, 1069 (1996), arXiv:hep-ph/9510338.

\bibitem{Tanabashi:1995nz}
M.~Tanabashi,
\newblock Phys.Lett. {\bf B384}, 218 (1996), arXiv:hep-ph/9511367.

\bibitem{Cornwall:1973tb}
J.~M. Cornwall, D.~N. Levin, and G.~Tiktopoulos,
\newblock Phys. Rev. Lett. {\bf 30}, 1268 (1973).
%%CITATION = PRLTA,30,1268;%%

\bibitem{Vayonakis:1976vz}
C.~E. Vayonakis,
\newblock Nuovo Cim. Lett. {\bf 17}, 383 (1976).
%%CITATION = NCLTA,17,383;%%

\bibitem{Chanowitz:1985hj}
M.~S. Chanowitz and M.~K. Gaillard,
\newblock Nucl. Phys. {\bf B261}, 379 (1985).
%%CITATION = NUPHA,B261,379;%%

\bibitem{Veltman:1976rt}
M.~J.~G. Veltman,
\newblock Acta Phys. Polon. {\bf B8}, 475 (1977).
%%CITATION = APPOA,B8,475;%%

\end{thebibliography}

\end{document}